\documentclass[pre,showpacs]{revtex4}

 \usepackage{amsmath,bm,amsthm,graphicx,enumerate}

 \newcommand{\Calbf}[1]{\pmb{\mathcal{#1}}}
 \newcommand{\Half}{\textstyle\frac{1}{2}}
 \newcommand{\Quarter}{\textstyle\frac{1}{4}}
 \newcommand{\D}{\textstyle{\rm d}}
 \newcommand{\E}{\textstyle{\rm e}}
 \newcommand{\I}{\textstyle{\rm i}}
 \newcommand{\bra}[1]{\mbox{$\langle#1\!\mid$}}
 \newcommand{\ket}[1]{\mbox{$\mid\!#1\rangle$}}

\begin{document}

 \title{Semi-spectral Chebyshev  method in Quantum Mechanics}

 \author{A. Deloff}
 \email{deloff@fuw.edu.pl}
 \affiliation{Institute for Nuclear Studies, Warsaw}

\date{\today}

 \begin{abstract}
 Traditionally, finite differences and finite element methods have been
 by many regarded as the basic tools for obtaining
 numerical solutions in a variety of quantum mechanical problems
 emerging in  atomic, nuclear and particle physics, astrophysics, 
 quantum chemistry, etc. 
 In recent years, however, an alternative technique based on
 the semi-spectral methods has focused considerable attention.
 The purpose of this work is first to provide the necessary tools 
 and subsequently examine the efficiency of 
 this method in quantum mechanical applications. Restricting
 our interest to time independent two-body problems, we obtained
 the continuous and discrete spectrum solutions of
 the underlying Schr\"{o}dinger or Lippmann-Schwinger 
 equations in both, the coordinate
 and momentum space.
 In all of the numerically studied examples we had no
 difficulty in achieving the machine accuracy
 and the semi-spectral method showed exponential 
 convergence  combined with  excellent numerical stability.

 \end{abstract}

 	 
\maketitle
 \section{Introduction}
 The exact solutions of the Schr\"odinger equation are known to exist
 only in a few rather idealized examples. Since in all remaining cases
 one has to be content with the numerically generated solutions,
 the importance of numerical methods in quantum mechanics
 can hardly be overestimated. Broadly speaking, 
 one may distinguish two schools of
 thought \cite{fornberg}\cite{boyd}:
 (i) the local approach comprising the finite differences
 and finite element methods, and, (ii) the global approach encompassing a
 variety of spectral methods. The local methods usually require much larger
 grids than other methods as the convergence of these schemes is
 generally algebraic. By contrast, the global methods have been shown to provide
 exponential convergence delivering smooth solutions
 which in most cases are preferable.
 \par
 The local methods approximate the unknown function by a sequence of
 overlapping low order polynomials which are used to interpolate the
 function at a small sub-set of the grid points. 
 The derivative of the unknown function is
 then approximated by the derivative of the interpolating polynomial so that
 the former takes the form of a weighted sum of the function values at
 the interpolation points. 
 The method is designed to be exact for polynomials of low order and 
 the form of the difference quotient is chosen such  
 that certain order of the truncation error is maintained (in the 
 lowest order $\D u(x)/\D x \approx [u(x+h)-u(x-h)]/(2h)+O(h^2)$
 where $h$ is a small grid spacing). In the simplest version this procedure 
 converts a differential equation into a three-term recurrence, 
 as may be exemplified by the popular Numerov algorithm \cite{taopang},   
 easy to program and implement but relatively low in accuracy.
 \par
 The strategy of the global or spectral methods is very different. 
 Here, the unknown function is sought
 in the form of a generalized Fourier expansion, truncated for practical reasons, 
 using global basis functions which usually are 
 polynomials or trigonometric polynomials of a high degree.
 This approach may be
 viewed as the limit of the finite difference method whose order of 
 accuracy has been pushed to infinity.
\par
The semi-spectral Chebyshev method 
-- a prominent representative of the global approach --
provides effective means to solve a generic
equation of the form $L\,u(x)=f(x)$ where the independent variable $x$ 
belongs to some interval $[a,b]$, $L$ is a differential or
integral operator, $f(x)$ is a given function and $u(x)$ is the unknown function
satisfying some, as yet unspecified boundary conditions. 
For the sake of simplicity of the exposition
we consider just a single independent variable $x$. The underlying idea is
exceedingly simple: to obtain the solution, first  
the problem is discretized by introducing a
grid consisting of $N$ points $t_i \in [a,b],\,i=1,2,\dots N$,
then the underlying equation is solved
{\em exactly} at all of the grid points, and, finally with the solutions
$u(t_i)$ in hand, an interpolating procedure is invoked to get $u(x)$ 
at an arbitrarily chosen $x$ value. 
Obviously, the key issue here would be the choice of
the grid and the interpolation method but it turns out that the scheme based on Chebyshev
polynomials provides the best offer on both accounts.  
It would be instructive to compare this approach with other methods
and put it in a historical perspective
but this is not so simple because the abundant mathematical literature
is lacking a generally accepted standardization \cite{boyd} and quite often
different methods bear the same tag, or else, the same thing appears
under different names. Accordingly, the semi-spectral method might be
called orthogonal collocation, Petrov-Galerkin, or even Rayleigh-Ritz method.
The common point of departure in all of the methods is the effort to minimize
the error distribution given by the reminder or the residue
function $R(x)\equiv L\,u(x)-f(x)$ and there
is a number of possible strategies to achieve this goal.
It is convenient to seek
the solution as a superposition of some {\em trial functions} $\phi_i(x)$
\begin{equation}
 u(x)=\sum_{i=1}^N c_i\,\phi_i(x),	
 \label{introd1}
\end{equation}
where $c_i$ are hitherto unknown generalized Fourier expansion coefficients.
The trial functions are expected to be easy to get by apart from the
usual indispensable requirements of completeness and orthogonality.
Local methods like finite difference or
finite element methods use for this purpose low order 
partially overlapping polynomials
defined on a grid whereas the global methods employ
trigonometric functions (Fourier methods) or high order polynomials
spanning over the whole domain. 
The global methods are also known as the spectral methods. In order to
minimize the residue function another set of the so called {\em test functions}
$\chi_i(x)$ is needed together with a definition of a scalar product
of two functions $f(x)$ and $g(x)$ which is taken to be given as 
\begin{equation}
	(f,g)=\int_a^b f(x)\,g(x)\,\rho(x)\,\D x,
 \label{introd2}
\end{equation}
where $\rho$ is a weighting function and the interval $[a,b]$ specifies the domain.
When $u$ were the exact solution the residue function would identically vanish 
 but with an approximate solution we can only try making  
the error distribution as small as possible by minimizing   
the residue function. This can be effected 
 by stipulating the conditions $(R,\chi_i)=0,\; i=1,2,\dots N$. Indeed,
 since the test functions form a complete set,
 the vector $R$ orthogonal to all members of such a
 set must be a null vector. The above orthogonality conditions 
 lead to  a linear set of algebraic equations
\begin{equation}
	\sum_{i=1}^N (\chi_j,L\phi_i)c_i=(\chi_j,f),\quad j=1,2,\dots N
 \label{introd3}
\end{equation}
 for the expansion coefficients $c_i$ 
 which can be solved by standard methods. 
 It will be convenient to distinguish three possible implementations
 of the above scheme.
 \begin{enumerate}[(i)]
	 \item {\it Semi-spectral method}. A grid $t_i$ is introduced and the test functions
		 are adopted in the form $\chi_i(x)=\delta(x-t_i)$.
	 \item  {\it Galerkin, or spectral method}. The test functions and the trial functions
		 are identical $\chi_i(x)=\phi_i(x)$ for all $i$.
	 \item  {\it Petrov-Galerkin method}. The test functions and the trial functions
		 are different.
 \end{enumerate}
 It has to be noted that the grid is indispensable 
 only in the semi-spectral method whereas the remaining
 two methods can do without it. In the following we shall concentrate
 on the semi-spectral and the Galerkin method. 
 \par
 As far as the basis functions are concerned, it
 is a common practice taking for  $\phi_j(x)$ 
 any of the sets of the orthogonal polynomials collected in \cite{abramowitz}. 
 Indeed, since all of them result as eigenfunctions of a certain
 Sturm-Liouville problem their completeness and orthogonality are guaranteed.
 There are also additional advantages following
 from taking polynomials as the expansion basis: firstly,
 this immediately sets an important correspondence 
 linking the generalized Fourier expansion with
 an interpolating scheme, and, secondly, it   
 simplifies all the necessary integrations by 
 using the highly accurate Gauss integration scheme.
 We shall take on both of these issues, assuming that from now on 
 $\phi_j(x)$ stands for a polynomial of the order $(j-1)$. 
 For an approximation of the order $N$ and assigned polynomial type
 the grid points $t_j,\;j=1,2,\dots N$ will be 
 identical with the Gauss abscissas which are given as the zeros of 
 the polynomial $\phi_{N+1}(x)$. 
 \par
 {\it Integration.} The Gauss integration formula, takes the form
 \begin{equation}
	 \int_a^b u(x)\,\rho(x)\,\D x =\displaystyle \sum_{j=1}^N w_j\,u(t_j)
 \label{introd4}
 \end{equation}
 and is known to be exact if $u(x)$ is a polynomial whose order
 is not greater than $(2\,N+1)$. The weights $w_j,\,j=1,2,\dots N$
 are obtained by solving the linear system
 \begin{equation}
 \sum_{j=1}^N t_j^n \, w_j=\int_a^b x^n\,\rho(x)\, \D x;\quad n=0,1,\dots (N-1). 
 \label{introd5}
 \end{equation}
 \par
 {\it Interpolation}. The truncated Fourier expansion
 can be converted into an interpolative formula with the aid of
 the {\em cardinal} function $G_j(x)$ with the property $G_j(t_i)=\delta_{ij}$.
 This function can be obtained by construction
 \begin{equation}
  G_j(x)=\phi_{N+1}(x)/[\phi^\prime_{N+1}(t_j)(x-t_j)]	 
 \label{introd6}
 \end{equation}
 where prime denotes the derivative and the interpolative formula reads
 \begin{equation}
	 u(x)=\sum_{j=1}^N \,u(t_j)\,G_j(x).
 \label{introd7}
 \end{equation}
 The cardinal function is a polynomial of order $(N-1)$ and may be written
 as a superposition of the basis functions
 \begin{equation}
	 G_j(x)=\sum_{i=1}^N M_{ji}^{-1}\,\phi_i(x)	 
 \label{introd8}
 \end{equation}
 where the coefficient matrix $\bm{M}^{-1}$ can be determined by taking advantage
 of the orthogonality of the basis. Using the Gauss integration formula, we obtain
 \begin{equation}
	 M^{-1}_{ij}=\phi_i(t_j)\,w_j /(\phi_i,\phi_i)
 \label{introd9}
 \end{equation}
 This matrix can be immediately inverted by applying Gauss integration 
 formula for expressing the orthogonality condition for the basis
 \begin{equation}
 \int_a^b\phi_ix)\,\phi_k(x)\,\rho(x)\,\D x=\delta_{ik}(\phi_i,\phi_i)=
 \displaystyle \sum_{j=1}^N \phi_i(t_j)\,w_j\,\phi_k(t_j)
 \label{introd10}
 \end{equation}
 and this result is exact. Since the orthogonality condition has here
 the form $\bm{M}^{-1}\cdot\bm{M}=\bm{1}$, it remains to read off the
 the matrix $\bm{M}$
 \begin{equation}
	 M_{ij}=\phi_j(t_i).
 \label{introd11}
 \end{equation}
 The function $u(x)$ is determined by supplying the coefficients $c_j$
 occurring in the spectral expansion formula but the latter is
 completely equivalent to the interpolative formula  
 that instead of the $c_j$ employs the grid values $u(t_j)$. There is
 a linear relation between these two arrays provided by the $\bm{M}$ matrix
 which can be written in a concise form
 \begin{equation}
	 [u] = \bm{M} \cdot [c].
 \label{introd12}
 \end{equation}
 The array of the coefficients $[c]$ is obtained by solving 
 the linear system (\ref{introd3}). In the semi-spectral method,
 this gives
 \begin{equation}
	 \sum_{j=1}^N \{L\phi_j(x)\}_{x=t_i}\,c_j=f(t_i),\quad i=1,2,\dots N
 \label{introd13}
 \end{equation}
 whereas the Galerkin scheme yields
 \begin{equation}
	 \sum_{j=1}^N (\phi_k,L\phi_j)\,c_j=(\phi_k,f),\quad k=1,2,\dots N
 \label{introd14}
 \end{equation}
 However, if the integration in (\ref{introd14}) is effected by the
 Gauss method, the resulting equations in both, semi-spectral and Galerkin schemes
 turn out to be identical and the two methods are completely equivalent.
 Indeed, multiplying both sides of (\ref{introd13}) by $\phi_k(t_i)\,w_i$
 and carrying out a summation over $i$, we immediately obtain (\ref{introd14}).
 In many applications the interpolative form (\ref{introd7}) might
 be more convenient, in which case $u(t_j)$ play the role of the unknowns.
 The latter are determined by solving the linear system
 \begin{equation}
	 \sum_{j=1}^N \{L\,G_j(x)\}_{x=t_i}\,u(t_j)=f(t_i),\quad i=1,2,\dots N
 \label{introd15}
 \end{equation}
 The resulting $u(t_j)$ 
 are subsequently used in (\ref{introd7}) to obtain the ultimate solution
 of the problem, as advertised at the beginning of this section.   
 \par
 In this paper we are going to use the Chebyshev polynomials as the basis
 functions. There are strong arguments corroborating this choice. Firstly, the
 Chebyshev polynomials optimize the interpolative procedure, as will be
 discussed shortly. Secondly, the values of the grid points $t_j$ 
 for all $N$ are given  
 by an elementary analytic expression whereas for other polynomials
 they would have to be computed numerically for every $N$. 
 Thirdly, the Fourier 
 coefficients of the function and the appropriate coefficients specifying the derivative
 are connected by a simple linear transformation. Fourthly, the same is
 true for the antiderivatives. Simplifying matters, one may say that by differentiating
 or integrating a set of Chebyshev polynomials we always obtain a set of Chebyshev
 polynomials and therefore these operations can be reduced to simple
 matrix manipulations.
 In the next Section we have presented all the necessary tools
 for solving some typical differential and integral equations occurring in
 quantum mechanics.  Subsequently, we consider the application of
 the semi-spectral Chebyshev method.
 Due to space limitations some selection was inevitable 
 and we decided to confine our interest to the
 time independent non-relativistic two-body problems.
 We concentrate our attention on the
 situation when the forces take the form of a short ranged potential
 considering in sequence the continuous and the discrete spectrum. 
 The solutions are obtained both in configuration and in momentum space
 using the appropriate Schr\"odinger, or the Lippmann-Schwinger
 (from now on referred to as L-S) equation,
 with and without the Coulomb interaction. These kind of problems
 are often encountered in nuclear and particle physics and we 
 offer a variety of complete solutions.
 \par
 In the last part we present some numerically solved examples.
 The short-range interaction is  modelled by three different 
 spherically symmetric potentials. The first of them, the exponential
 potential can be viewed as the simplest prototype of the forces
 rapidly falling off with the distance \cite{morsefeshbach}. 
 More complicated shapes
 can be obtained by taking suitable superpositions of exponential
 potentials with different ranges. Our second choice, 
 the Morse potential \cite{morse}, is a particular combination of 
 two exponential potentials of different ranges where one of
 them is repulsive while the other is attractive. This potential
 has been very popular \cite{rawitscher02} 
 as a model for vibrational states of diatomic molecules
 but it has been also used in nuclear physics
 to simulate a repulsive core nucleon-nucleon interaction \cite{darewychgreen}.
 As the other extreme one may take  
 an infinite superposition of exponential potentials forming 
 a geometric series. The explicit summation of this series leads to 
  our third choice, the familiar
  Hulth\'{e}n potential \cite{hulthen} which close to the origin develops a $1/r$ singularity,
 like the Coulomb potential, and yet exhibits an exponential fall-off at
 large distances.
 The meaning of the Hulth\'{e}n potential goes far beyond academic interest
 as it has been widely used in nuclear and particle phenomenology 
 \cite{durand}\cite{vandijk}, 
 atomic physics solid state \cite{lamvarshni}, quantum chemistry \cite{olsonmicha} etc.
 Here, for us the important fact is that for $L=0$ all the three potentials
 mentioned above admit analytic solutions and the situation
 has been summarized in the Appendix \cite{morsefeshbach}\cite{fluge}.
 In most instances the semi-spectral method
 turns out to be so accurate that the standard procedure,
 where the convergence is examined
 by comparison with the results obtained by ''other methods'',
 almost certainly would be meaningless.
 Therefore, for calculating the relative errors
 we  always utilized the exact results.
\section{Chebyshev interpolation}
\subsection{Chebyshev cardinal function}
 Let us consider an independent variable $t$ defined in the $[-1,1]$ interval
 (our conventions and notation in this section are consistent with \cite{recipes})
The Chebyshev polynomial $T_N(t)$ of the first kind and of degree $N$ is defined 
\cite{mason}\cite{rivlin} by the formula
 \begin{equation}
T_N(t)=\cos(N \arccos(t)).
 \label{G0}
 \end{equation}
 It is evident that $T_N(t)$  has $N$ zeros in the $[-1,1]$ interval
 and they are located at the points
\begin{equation}
t_k=\cos[\pi(k-\Half)/N]; \quad k=1,2,\dots, N
\label{G1}
\end{equation}
The Chebyshev polynomials can be also generated from the recurrence relation
\begin{equation*}
        T_n(t)=2t\,T_{n-1}(t)-T_{n-2}(t)
\end{equation*}
with $T_0(t)=1$ and $T_1(t)=t$.
The Chebyshev polynomials of the second kind, denoted as $U_n(t)$, are obtained
from the same recurrence 
relation but with different starting values: $U_{-1}(t)=0$ and $U_0(t)=1$.
\par
 Usually, two choices of grids
 are considered: (i)  the classical Chebyshev mesh-points 
 given in (\ref{G1}), or, (ii) the zeros of the derivatives of $T_N(t)$ which is the same
as the zeros of $U_{N-1}(t)$.  The choice (ii) is called  Lobatto mesh, for it was shown by
Lobatto that the zeros of the derivative $T'_N(t)$ supplemented by the end-points $\pm 1$ 
have all the advantageous features of the Chebyshev mesh.  Furthermore, if  a function
is approximated by Chebyshev polynomials  supplying the boundary values of the function
immediately pins down two coefficients.  The other attractive feature shows up when the
Lobatto mesh is used in quadratures and when $N$ and subsequently $2N$ points
are taken 
to obtain useful and computable error estimate. In the bigger mesh half of the points
coincide with those of the smaller mesh and therefore the function values calculated
at the smaller mesh can be later reused reducing thereby the total number of function
evaluation. The Lobatto mesh, however, is not  always preferable and especially  
for functions which exhibit end-point singularities more convenient 
might be the classical Chebyshev mesh.
Apart from personal prejudices,
there is little difference between these two schemes in terms of accuracy,
or convergence \cite{boyd}.
\par    
The Chebyshev polynomials are orthogonal in the interval $[-1,1]$ over 
a weighting factor $1/\sqrt{1-t^2}$ and from the continuous orthogonality relation 
 a discrete orthogonality relation can be derived.
In this work we shall use the classical Chebyshev mesh (\ref{G1})
and the appropriate orthogonality relations, take the form            
\begin{equation}
\dfrac{\pi}{N}\sum_{k=1}^N T_i(t_k)\,T_j(t_k)=
\delta_{ij}\;\dfrac{\pi}{2}(1+\delta_{0i})
\label{G1a}
\end{equation}
where $i<N$ and $j<N$.  
\par 
A continuous and bounded function $f(t)$ can be approximated  in $[-1,1]$ interval
by the expression
\begin{equation}
f(t)=\sum_{j=1}^N {\rm '\mbox{} } \ c_j\,T_{j-1}(t)
\label{G2}
\end{equation} 
where the primed sigma denotes hereafter a summation in which the first term should be halved.
 The spectral coefficients $c_j$ can be obtained 
by multiplying (\ref{G2}) by $T_{i-1}(t)$ and then  setting $t$ equal to $t_k$
and, finally, performing a summation over $k$. Making use of (\ref{G1a}), we get
\begin{equation}
	c_j= \dfrac{2}{N}\sum_{k=1}^N \, M_{jk}\, f(t_k),
\label{G3}
\end{equation} 
 with $M_{jk}=T_{j-1}(t_k)$ where 
\begin{equation}
 T_{j-1}(t_k) = \cos[\pi(k-\Half)(j-1)/N].
\label{G4}
\end{equation}
 We may regard (\ref{G2}) as a generalized truncated Fourier expansion.  
 Recalling that $T_N(t)$ vanishes at all $t=t_j$, (\ref{G2}) can be rewritten as
\begin{equation}
        f(t) =  \sum_{j=1}^N\dfrac{T_N(t)}{T_N^\prime(t_j)(t-t_j)}\;f(t_j),
        \label{G2a}
\end{equation}
where prime stands for the derivative. This is nothing else but
an interpolating formula, which may be cast to the familiar Lagrange form
\begin{equation}
        f(t) =  \sum_{j=1}^N \;G_j(t)\;f(t_j),
        \label{G2b}
\end{equation}
where we have introduced the Chebyshev
cardinal functions $G_j(t)$, with the property $G_j(t_i)=\delta_{ij}$.
 Indeed, inserting  (\ref{G3})  into  (\ref{G2}), we obtain
\begin{equation}
        G_j(t)=\dfrac{2}{N}\sum_{i=1}^N {\rm '\mbox{} } \ T_{i-1}(t_j)\,T_{i-1}(t).
        \label{G2c}
\end{equation}   
  It is evident that there are two options for reconstructing
  the function $f(t)$:  either from the expansion (\ref{G2}) knowing the
  coefficients $c_i$, or by
  supplying the values of $f(t_j)$ and using (\ref{G2b}). 
  These options
  are completely equivalent and 
  knowing the mesh-point values $f(t_j)$ of the function,
  the corresponding Fourier coefficients can be obtained from (\ref{G3}).
\par
 Similar procedure may be applied for a function of two, or more,
 variables. Thus, a simple extension of (\ref{G2b}) to 
 the case when the function depends upon two variables
 $x$ and $y$, is
\begin{equation}
        f(x,y) =        \sum_{i,j=1}^N \;G_i(x)\;f(t_i,t_j)\;G_j(y).
        \label{G2B}
\end{equation}
\par
Owing to their rapid, often exponential, convergence rate when
the number of mesh-points $N$ increases, the Chebyshev interpolation methods
have proved to be extremely efficient in solving various problems
of mathematical physics. This success has a very solid background  
resting on two celebrated theorems which we quote here without proof \cite{boyd}. 
\newtheorem{th1} {Theorem}
\begin{th1}[Cauchy interpolation error theorem]
  Let $f(t)$ be a function sufficiently smooth so that it has   
 at least  $N+1$  continuous derivatives on the interval $[-1,1]$
and let $P_N(t)$ be its Lagrangian interpolant of degree $N$ for $t\in[-1,1]$. 
 Then, for any grid ${t_i}\in[-1,1]$ and any $t\in[-1,1]$ 
 the upper bound of the interpolation error is given by
\begin{equation}
f(t)-P_N(t)\leq \dfrac{1}{(N+1)!}\;f^{(N+1)}(\xi)\;\prod_{i=1}^N(t-t_i)
\label{Th1}
\end{equation}
for some $\xi(t)\in[a,b]$.  
\end{th1}
It is worth noting that the product term  in (\ref{Th1}) is itself a monic polynomial 
   (leading term coefficient  equal unity). 
The above Cauchy theorem indicates that 
without specifying the function $f(t)$, the only way to minimize 
the interpolation error is
to minimize the product term in (\ref{Th1}) and this in turn can be
achieved by a judicious choice of the interpolation points. 
The answer how to accomplish that brings the following Chebyshev theorem \cite{boyd}.
\newtheorem{th2}[th1] {Theorem}
\begin{th2}[Chebyshev minimal amplitude theorem]
 Out of all monic polynomials of degree $N$,  the unique polynomial which has the
smallest maximum on $[-1,1]$ is the Chebyshev polynomial $T_N(t)$ divided by
$2^{N-1}$, i.e.  all monic polynomials $Q_N(t)$  of degree $N$ satisfy the inequality
\begin{equation}
\max |Q_N(t)|\geq \max |T_N(t)/2^{N-1}|=1/2^{N-1}
\label{Th2}
\end{equation}
for all $t\in[-1,1]$.
\end{th2}
\subsection{Anti-derivative}
 The optimized interpolation is highly satisfactory but there is much more to come. 
As we shall see in a moment, by using  
the spectral representation
 simple operations like differentiation or integration 
 can be reduced to matrix multiplication. We start with integration, introducing
 the anti-derivative of $f(t)$, defined as
\begin{equation} 
F^-(t)=\int_{-1}^t\, f(x)\,\D x.
\label{G5}
\end{equation} 
  The anti-derivative 
  might be sought in the form of an appropriate Chebyshev expansion
\begin{equation}
F^-(t)=\sum_{j=1}^N {\rm '\mbox{} } \ C_j\,T_{j-1}(t).
\label{G6}
\end{equation}
 in which case
 our objective would be to calculate the coefficients $C_j$ occurring in (\ref{G6}).
 If the function in the integrand  of (\ref{G5}) is represented by its Chebyshev expansion 
 (\ref{G2}), 
 the integration in (\ref{G5})  
 can be done analytically (Curtis-Clenshaw integration \cite{clenshaw}).
 This solves the problem because the coefficients  
 $C_j$ in (\ref{G6}) must be expressible in terms of $c_j$  
 but these relations are not particularly simple.
 El-gendi observed \cite{elgendi} that much simpler relations would be obtained
 if instead of the spectral coefficients the function values at the mesh-points were used
 \cite{elgendi}\cite{mihaila}. 
 In the latter case the array of the mesh values of the anti-derivative $F^-(t_i)$  would  be connected 
 with the array of the integrand mesh values $f(t_j)$ by a simple linear
 transformation which bears a universal character being independent upon the shape of $f(t)$.
 Indeed, by inserting (\ref{G2b}) in (\ref{G5}), we end up with
 a remarkably simple quadrature rule 
\begin{equation}
F^-(t_i)=\sum_{j=1}^N  \ W^-_{ij}\,f(t_j)
\label{G7}
\end{equation}
 where  $W^-_{ij}$ may be viewed as a generalized weighting factor.
 For a fixed  $N$,  this is a {\it constant} matrix, {\it viz.}  
\begin{equation}
        W^-_{ij}=\int_{-1}^{t_i}\, G_j(t)\,\D t
\label{G8}
\end{equation}
 which will be obtained in an analytic form.
 Inserting (\ref{G2c}) in (\ref{G8}), we get
\begin{equation}
  W^-_{ij}= \dfrac{2}{N}\sum_{k=1}^N {\rm '\mbox{} } \
  \int_{-1}^{t_i}\,T_{k-1}(t)\,\D t
 \;  T_{k-1}(t_j)
\label{G9}
\end{equation}
 and,
 as seen from (\ref{G9}), the matrix $\bm{W}^-$ may be written as a product
 $\bm{W}^-=\bm{S}^-\cdot \bm{d}^{-1} \cdot \bm{M}$, where
 $\bm{M}$ is given in (\ref{G4}), 
 $ \bm{d} = \text{diag}(N,\,\Half N,\, \Half N,\,\dots,\, \Half N) $, and
 $\bm{S}^-$, is 
\begin{equation}
        S^-_{ij} = \int_{-1}^{t_i}\,T_{j-1}(t)\,\D t.
        \label{G10a}    
\end{equation} 
  The integral in (\ref{G10a}) can be obtained from an analytic expression 
\begin{equation}
\int_{-1}^{t_i}\,T_{j-1}(t)\,\D t=
\begin{cases}
T_1(t_i)+1;       \quad &\text{for $j=1$}\\
[T_2(t_i)-1]/4; \quad &\text{for $j=2$}\\
T_j(t_i)/(2j)-T_{j-2}(t_i)/2(j-2)+(-1)^j/j(j-2); \quad & \text{for $j>2$}
\end{cases}
\label{G10}
\end{equation} 
 where the right hand sides of (\ref{G10}) are readily computed with
 the aid of (\ref{G4}). It is worth noting that
the discrete orthogonality relation (\ref{G1a}) can be written
in matrix form as
       $ \bm{M}\cdot\tilde{\bm{M}}=\bm{d} $
 with  tilde denoting the transposed matrix. 
 Since the matrix $\bm{M}$ has orthogonal columns, we have
 $\bm{M}^{-1}= \tilde{\bm{M}} \cdot \bm{d}^{-1}$.
 When the array $F^-(t_i)$ has been determined, the expansion coefficients
 of the anti-derivative may computed from
\begin{equation}
        C_i= \dfrac{2}{N}\sum_{j=1}^N \, M_{ij}\, F^-(t_j)
\label{G11}
\end{equation} 
 and the anti-derivative $F^-(t)$  could be reconstructed 
 at any point either from (\ref{G6}), or from
 (\ref{G2b}) after making the substitution $f\to F^-$.
\par
 Alternatively, we could have defined the anti-derivative as 
\begin{equation}
 F^+(t) = -\int_t^1\, f(x)\,\D x  
\label{G16}
\end{equation} 
 in which case the appropriate integration rule, takes the form    
\begin{equation}
F^+(t_i)=-\sum_{j=1}^N  \ W^+_{ij}\,f(t_j)
\label{G17}
\end{equation}
where $\bm{W}^+=\bm{S}^+\cdot\bm{d}^{-1}\cdot\bm{M}$ and for $\bm{S}^+$,
we obtain
\begin{equation}
\int_{t_i}^{1}\,T_{j-1}(t)\,\D t=
\begin{cases}
-T_1(t_i)+1;       \quad &\text{for $j=1$}\\
-[T_2(t_i)-1]/4;   \quad &\text{for $j=2$}\\
-T_j(t_i)/(2j)+T_{j-2}(t_i)/2(j-2)-1/j(j-2); \quad & \text{for $j>2$}
\end{cases}
\label{G18}
\end{equation} 
In the following we shall need both forms of the anti-derivatives.  
\subsection{Integration}
 A definite integral in the $[-1,1]$ limits might be calculated in a similar way
 as the anti-derivative but
we wish to consider the numerical evaluation of a slightly more general integral
\begin{equation}
 I[f]= \int_{-1}^1\,k(t)\, f(t)\,\D t
\label{G12a} \end{equation} 
where $k(t)$ is a real absolutely integrable function, which need not be continuous
or of one sign, while $f(t)$ is any continuous function.  The integration will be performed
by using the product integration form
\begin{equation}
 I_N[f]= \sum_{i=1}^N\,w_i\,f(t_i)
\label{G12b} \end{equation}  
where the weights $w_i$ are determined by requiring the above rule to
be exact when $f(t)$ is any polynomial of degree  $\leq N$.  This can be 
regarded as a variation of the Clenshaw-Curtis integration and indeed reduces to that
method when the mesh-points are taken to be the Lobatto points an $k(t)$ is set to unity.
It has been shown \cite{sloan} that the integration rule (\ref{G12b}) 
for $N\to\infty$ converges to
the exact result for all continuous functions $f(t)$, provided $k(t)$ satisfies a
rather mild integrability condition, namely
\begin{equation}
  \int_{-1}^1\,|k(t)|^p\,\D t<\infty
\label{G12c}
\end{equation}
for some $p>1$.  Two choices of $k(t)$ are of particular interest. The first is
when $k(t)$ is taken to be the Chebyshev weighting function $k(t)\equiv 1/\sqrt{1-t^2}$
and we end up with the standard Gauss-Chebyshev quadrature \cite{abramowitz}
\begin{equation}
\int_{-1}^1\, f(t)\,\dfrac{\D t}{\sqrt{1-t^2}}=
\dfrac{\pi}{N}\displaystyle\sum_{j=1}^N \,f(t_j),
\label{G12d}
\end{equation}
with   $w_j=\pi/N$ for all $j$. 
The second case of practical interest is that  in which  $k(t)\equiv 1$ and we have
\begin{equation}
\int_{-1}^1\, f(t)\,\D t=\displaystyle\sum_{j=1}^N \, w_j\,f(t_j)
\label{G12}
\end{equation} 
with the Chebyshev weights 
\begin{equation}
w_j= \dfrac{2}{N}\sum_{k=1}^N {\rm '\mbox{} } \ T_{k-1}(t_j)\int_{-1}^1 \,T_{k-1}(t)\,\D t.
\label{G13}
\end{equation}
The integral occurring in (\ref{G13}) can be easily evaluated, {\it viz.}   
\begin{equation}
\int_{-1}^1 \,T_{k-1}(t)\,\D t=
  [(-1)^k-1]/ [k(k-2)]
\label{G14}
\end{equation} 
 and inserting (\ref{G14}) in (\ref{G13}), we arrive at  the ultimate expression
for the weights
\begin{equation}
 w_j= -\dfrac{4}{N}\;\;\sideset{}{ \,'}\sum_{i=0}^{[(N-1)/2]} \;\;
        \dfrac{ T_{2i}(t_j) }
     { 4i^2-1 }.
\label{G15}
\end{equation}
 It can be proved that
 the weights (\ref{G15}) are all positive and 
 setting $f(t)\equiv 1$ in (\ref{G12}), we obtain the sum rule
 \begin{equation*}
         \sum_{i=1}^N w_i = 2.
 \end{equation*}
 \subsection{Singular integrals}
 The Chebyshev quadrature can be extended to comprise also singular integrals.
 To this end we consider product integration (\ref{G12a}) in which 
 $k(t)$  may contain isolated singularities \cite{hasegawa}.
 Perhaps the most important of them is the Cauchy integral, 
 and we wish to establish the quadrature rule for the principal value
 integral of the form
 \begin{equation}
         \int_{-1}^1 \dfrac{f(t)}{t-z}\,\D t =\displaystyle \sum_{j=1}^N\omega_j(z)\,f(t_j)
         \label{sing1}
 \end{equation}
 where the weights $\omega_j(z)$ depend upon the location of the singularity. 
 It is
 assumed in the following that $|z|<1$ and that the function $f(t)$ is free from singularities
 on the $<-1,1>$ interval of the real axis. Our goal is to calculate the generalized weights
 $\omega_j(z)$ and inserting (\ref{G2b}) in the left hand side of (\ref{sing1}),
 we have 
 \begin{equation}
         \omega_j(z)
         = \dfrac{2}{N}\sum_{k=1}^N {\rm '\mbox{} } \;
         T_{k-1}(t_j) \,I_{k-1}(z)
         \label{sing2}
 \end{equation}
 with $I_{k-1}(z)$ denoting the principal value integral
 \begin{equation}
 I_{k-1}(z)=
 \int_{-1}^1 \dfrac{T_{k-1}(t)}{t-z}\,\D t 
         \label{sing3}
 \end{equation}
 which will be calculated in an analytic form. Indeed, a simple subtraction, yields   
 \begin{equation}
 \int_{-1}^1 \dfrac{T_{k-1}(t)}{t-z}\,\D t= \displaystyle
 \int_{-1}^1 \dfrac{T_{k-1}(t)-T_{k-1}(z)}{t-z}\,\D t+T_{k-1}(z)\,\log{\dfrac{1-z}{1+z}} 
         \label{sing4}
 \end{equation}
 and the integrand on the right hand side of (\ref{sing4}) being
 now regular, can be represented as a superposition of Chebyshev
 polynomials
 \begin{equation}
  \dfrac{T_{k-1}(t)-T_{k-1}(z)}{t-z}
  = 2\sum_{i=0}^{k-2} {\rm '\mbox{} } \;U_{k-2-i}(z)\,T_i(t) 
         \label{sing5}
 \end{equation}
 rendering the t-integration straightforward. With the aid of (\ref{G14}), we obtain the  
  ultimate  expression
 \begin{equation}
 I_{k-1}(z)=
 S_{k-1}(z) +T_{k-1}(z)\,\log{[(1-z)/(1+z)]};  
         \label{sing5a}
 \end{equation}
 with
 \begin{equation}
 S_{k-1}(z) \equiv 
 -2\displaystyle\sum_{i=0}^{k-2}{\rm'\mbox{} }\;\dfrac{ (-1)^i+1 }
 {i^2-1}\; U_{k-2-i}(z)
         \label{sing6}
 \end{equation}
 and (\ref{sing5a})-(\ref{sing6}) 
 used in (\ref{sing2}) allows to calculate explicitly the weights. 
 The presence of
 the end-point logarithmic singularities is apparent
 in (\ref{sing5a}) and at these values the principal value integral
 (\ref{sing3}) becomes undefined. It is interesting to note
 that the relation between  the function
 $I_n(z)$ and $T_n(z)$ is analogous to that between
  the Legendre function $Q_n(z)$ and the Legendre polynomial $P_n(z)$.
  It will be convenient to separate 
  the singular term writing  the weights in the form
  \begin{equation}
          \omega_j(z)=\tilde{\omega}_j(z)+G_j(z) \,\log{[(1-z)/(1+z)]};  
          \label{sing6a}
  \end{equation}
  where $G_j(z)$ is the cardinal function and the regular part, given by 
 \begin{equation}
         \tilde{\omega}_j(z)
         = \dfrac{2}{N}\sum_{k=1}^N {\rm '\mbox{} } \;
         T_{k-1}(t_j) \,S_{k-1}(z)
         \label{sing6b}
 \end{equation}
 is just a polynomial in $z$.
 The ultimate quadrature rule for a general integral of the Cauchy type, reads
 \begin{equation}
         \int_{-1}^1 \dfrac{f(t)\,\D t}{t-z\mp\I\epsilon } =
 \begin{cases}
         \displaystyle \sum_{j=1}^N \dfrac{w_j} {t_j-z}\,f(t_j), &  |z|>1 \\ 
         \displaystyle \sum_{j=1}^N \tilde{\omega}_j(z)\,f(t_j)
     +f(z)\left( \log{\dfrac{1-z}{1+z}} \pm\I\pi\right). & |z|<1
 \end{cases}
         \label{sing6c}
 \end{equation}
 \par
 Setting $f(t) \equiv 1$ in (\ref{sing1}), a sum rule is obtained
 \begin{displaymath}
 \displaystyle \sum_{j=1}^N\omega_j(z)=
 \log{\dfrac{1-z}{1+z}} 
 \end{displaymath}
  providing a convenient check of the computed weights. 
 \par
 The other important case to be considered here is that of the logarithmic singularity.
 The appropriate automatic quadrature, takes the form 
 \begin{equation}
         \int_{-1}^1 f(t)\, \log{|t-z|}\,\D t =
 \begin{cases}
         \displaystyle \sum_{j=1}^N w_j \,f(t_j)\log{|t_j-z|},  & |z|>1 \\
         \displaystyle \sum_{j=1}^N\Omega_j(z)\,f(t_j),         & |z| \leq 1
 \end{cases}
         \label{sing13}
 \end{equation}
 and our goal is to determine the weights $\Omega_j(z)$.
 The assumptions concerning $f(t)$ and $z$ remain the same as in
 the Cauchy integral case (\ref{sing1}). In particular, for the time being, we assume
 that $|z| \neq 1$ but eventually this restriction will be lifted.
 Upon inserting (\ref{G2b}) in the left hand side of (\ref{sing13}), 
 the  dedicated weights $\Omega_j(z)$ can be written in the form
 \begin{equation}
         \Omega_j(z)
         = \dfrac{2}{N}\sum_{k=1}^N {\rm '\mbox{} } \;
         T_{k-1}(t_j) \,J_{k-1}(z)
         \label{sing14}
 \end{equation}
 with $J_{k-1}(z)$ denoting the integral
 \begin{equation}
 J_{k-1}(z)=
 \int_{-1}^1 T_{k-1}(t)\;\log{|t-z|}\,\D t, 
         \label{sing15}
 \end{equation}
 which can be evaluated  analytically. In order to do that,
 we take advantage of the identity \cite{rivlin} which holds for non-negative $n$
 \begin{displaymath}
         T_n(t)=
 \begin{cases}
         \dfrac{1}{2(n+1)}\,T_{n+1}^\prime(t)    
         -\dfrac{1}{2(n-1)}\,T_{|n-1|}^\prime(t),\quad & n\neq 1 \\
         \frac{1}{4}\,T_2^\prime (z), &  n=1.
 \end{cases}
 \end{displaymath}
 with prime denoting the derivative. 
 Inserting the above identity in (\ref{sing15}), 
 the integration by parts, gives
 \begin{equation}
 J_{k-1}(z)=
 \begin{cases}
 -\dfrac{\log{|1-z|}-(-1)^k\log{|1+z|}}{k(k-2)}
 -\dfrac{I_k(z)}{2k}
 +\dfrac{I_{|k-2|}(z)}{2(k-2)}, & k\neq 2\\
 \frac{1}{4}
 \left\{\log{\left|\dfrac{1-z}{1+z}\right|}-I_2(z)\right\},& k=2,
 \end{cases}
         \label{sing16}
 \end{equation}
 where $k=1,2,\cdots, N$.
 Substituting the explicit forms of $J_{k-1}(z)$ 
 in (\ref{sing14}) completes the derivation
 of the weighting functions $\Omega_j(z)$. 
 \par
 The integral
 containing a logarithmic singularity (\ref{sing13}), unlike the Cauchy integral,
 does exist even when the singularity coincides with either of the 
 integration end points. Therefore, all $J_{k-1}(z)$ in (\ref{sing16}) go, as they should,  
 to a finite limit when $z\to\pm 1$
 and the explicit expressions, are
 \begin{equation}
 J_{k-1}(\pm 1)=
 \begin{cases}
         2\log 2 -2,   & k=1\\
         \mp 1, &  k=2\\
         -\dfrac{[1+(-1)^{k-1}]\,\log 2}{k(k-2)}-\dfrac{1}{2k}\,S_k(\pm 1)
         +\dfrac{1}{2(k-2)}\,S_{k-2}(\pm 1), &  k\ge 3 
  \end{cases}
         \label{sing16a}
 \end{equation}
 where  $S_k(\pm 1)$ and $S_{k-2}(\pm 1)$ are computed from (\ref{sing6}).
 With (\ref{sing16a}) in hand, we can evaluate $\Omega_j(\pm 1)$ from (\ref{sing14}),  
 establishing the automatic quadrature (\ref{sing13}) for the case when the singularity
 is located at the integration end points.
 \par
 Setting $f(t)\equiv 1$ in (\ref{sing13}),
 we arrive at the sum rule 
 \begin{displaymath}
         \sum_{j=1}^N\,\Omega_j(z)=J_0(z)=
         (1-z)\log{(1-z)} + (1+z)\log{(1+z)} - 2
 \end{displaymath}
 which might be useful for checking purposes. 
 The automatic quadrature just presented has usually a very
 high rate of convergence and compares favorably with
  a direct method of calculating the integral (\ref{sing13}) 
 where the singularity is eliminated by  subtraction. 
   \subsection{Differentiation}
 The Chebyshev expansion is also well suited for performing
 differentiation. In fact, we could invert our formulae for
 integration to derive the appropriate expressions for 
 differentiation. It is better, however, to provide an
 {\it ab initio} derivation. Upon differentiating (\ref{G2}), we have 
\begin{equation}
  \dfrac{\D f(t)}{\D t}=\sum_{j=1}^N {\rm '\mbox{} } \ c_j\,\dfrac{\D T_{j-1}(t)}{\D t}
\label{GA1}
\end{equation} 
but this function has also a standard Chebyshev expansion
\begin{equation}
        \dfrac{\D f(t)}{\D t}\equiv f'(t_)
        =\sum_{j=1}^N {\rm '\mbox{} } \ c_j^{\prime}\, T_{j-1}(t)
\label{GA2}
\end{equation}
where $c^{\prime}_j$ denote the appropriate
 Fourier-Chebyshev expansion coefficients.
 Clearly, $c^{\prime}_j$ denote the appropriate
are related to the expansion coefficients for the function $c_j$. The same is true
for the  $f'(t_i)$ and $f(t_i)$ and there is a linear transformation connecting
these two arrays. To find the transformation matrix explicitly we set in (\ref{GA1}) $t=t_i$ 
inserting  for $c_j$ formula (\ref{G3}). As a result, we obtain  
\begin{equation}
f^\prime(t_i)=\sum_{j=1}^N  \ D_{ij}\,f(t_j).
\label{GA3}
\end{equation}
where the matrix $D_{ij}$ is
\begin{equation}
  D_{ij}= \dfrac{2}{N}\sum_{k=1}^N {\rm '\mbox{} } \
  \,T_{k-1}^\prime(t_i)
 \;  T_{k-1}(t_j).
\label{GA4}
\end{equation}
The derivative of the Chebyshev polynomial occurring in (\ref{GA4}) can be  
can be easily computed as
\begin{equation}
        T_{k-1}^\prime(t_i)=(k-1) \, U_{k-2}(t_j)
        \label{GA5}
\end{equation}
 where $U_{k-2}(t_j)$  denotes a Chebyshev polynomial of the second kind. 
\par
 When the array $f'(t_i)$ is known, the expansion coefficients are computed from
\begin{equation}
        c^\prime_i= \dfrac{2}{N}\sum_{j=1}^N \, M_{ij}\, f'(t_j)
\label{GA7}
\end{equation} 
 and the function $f'(t)$ can obtained  form (\ref{GA2}).  
 \subsection{ Arbitrary interval $[a,b]$ } 
 In the above considerations the independent variable $t$ was confined to the $[-1,1]$
 interval.  This restriction may be lifted by introducing a new independent variable
 $x$ defined in an arbitrary interval $[a,b]$. 
 A linear mapping generates the appropriate Chebyshev mesh in $[a,b]$
 \begin{equation}
         x_j=\Half(b+a)+\Half(b-a)\;t_j;\quad j=1,2,3,\dots,N
         \label{G20}
 \end{equation}
 The cardinal functions (\ref{G2c}) constituting the basis
 for the interpolation formula 
 depend now upon the argument obtained by the inverse transformation.
 The extension of (\ref{G2b}), is
\begin{equation}
        f(x) =  \sum_{j=1}^N \;G_j\left(\dfrac{2x-a-b}{b-a}\right)\;f(x_j)
        \label{G20a}
\end{equation}
 and the Gauss-Chebyshev quadrature reads
 \begin{equation}
         \int_a^b f(x) \D x = \Half (b-a)\displaystyle\sum_{j=1}^N w_j f(x_j) 
         \label{G21}
 \end{equation}
 with the weights given in (\ref{G15}). The modifications for other automatic integrations
 are rather obvious.
 \par
 The anti-derivative is obtained
 in a similar way: the weights acquire the $\Half(b-a)$ factor, 
 induced by the change of variable, and the
 integrand must be evaluated at the  mesh-points (\ref{G20}). 
 Thus, the extension of  (\ref{G7}), is
\begin{equation}
        F^-(x_i)= \int_a^{x_i} f(x)\,\D x =
 \Half(b-a)     \displaystyle\sum_{j=1}^N  \ W^-_{ij}\,f(x_j).
\label{G22}
\end{equation}
where the weighting matrix $\bm{W}^-$  once and for all is given in (\ref{G9}).
\par
In order to apply the Chebyshev method 
for calculating semi-definite integrals there are two possibilities: (i) truncation followed by
a linear mapping, or (ii) non-linear mapping.
Truncation method replaces the infinite integration 
domain $[0,\infty]$ by a  large albeit finite  
domain $[0,R]$. Non-linear mapping consists in 
 a change of variables devised in such a way 
that the original integral is converted into an integral which is in the $[-1,1]$ limits. 
Clearly, there would be a countless number
of such transformations but the simplest one is the rational mapping
\begin{equation}
r=R(1+t)/(1-t),
\label{G22a} 
\end{equation}
which relates the original variable $r\in[0,\infty]$ with  a new variable $t\in[-1,1]$.
Dealing with improper integrals has its price:
in both cases an adjustable parameter $R$ appears, in one case as a cut-off
in the other (\ref{G22a}) as a scaling, or a slope parameter,
 whose value could be optimized at
the expense of some trial-and-error procedure. 
 In general, the truncation method is more
sensitive to the variation of $R$. Actually, the two numerical parameters $N$ and $R$
are correlated and some experimentation is inevitable if one wants to minimize the error    
 by  increasing  them simultaneously.  
In addition to the appearance of the truncation or scaling parameter $R$
the infinite integration range induces also another unwelcome feature 
in that the increase of $R$ brings  all the poles and branch point singularities
of the integrand (or its derivatives) closer to the real axis after  the original variable has been
rescaled to the $[-1,1]$ interval.  The presence of singularities  close to the
integration region in consequence leads to a deterioration of the convergence rate.  
 \section{ Applications }               
The implementation of the semi-spectral Chebyshev method is very simple because
 all that is needed is a standard linear system solver plus a set of procedures
calculating the cardinal functions $G_j(t)$, the anti-derivative matrices $\bm{W}^\pm$,
the differentiation matrix $\bm{D}$ and the dedicated weights $w_j, \omega_j(z), \Omega_j(z)$.
 Explicit means for getting all of them have been described in detail in Sec. II.
Assuming then that the toolbox specified above is available, we are going to review
some applications of the semi-spectral method that are of relevance to
quantum mechanics. The scheme remains the same in all problems: 
the equation to be solved is first discretized and on the grid
the integration and differentiation operators are represented by appropriate matrices. 
As a result, the mesh-point values of the sought for function 
can be pined down by solving a linear system of algebraic equations. Finally,
by feeding the interpolation formula (\ref{G2b}) with the mesh-point function values, 
  the solution at any point can be obtained.
\subsection{Second order differential equation}
 The second order differential equation of 
 considerable interest in physics (Schr\"{o}dinger equation), has the form
 \begin{equation}
         y''(x)+p(x)\,y(x)=q(x);\quad a\leq x\leq b
         \label{G46}
 \end{equation}
 where $p(x)$ and $q(x)$ are given functions. The unknown function $y(x)$
 and its derivative are assumed to satisfy some boundary conditions
 at $a$ or/and at $b$ depending on the nature of the  physical problem at issue.
 We consider first the case when the values of 
 $y(a)$ and $y'(a)$ have been assigned.
 When (\ref{G46}) is integrated twice from $a$ to $x$,  we obtain
 \begin{equation}
         y(x)-y(a) - y'(a)(x-a)+\int_a^x\D s\displaystyle\int_a^s \D t\, p(t)\,y(t)  
         =\displaystyle\int_a^x\D s\displaystyle\int_a^s \D t\, q(t).  
         \label{G47}
 \end{equation}
 Introducing the Chebyshev mesh (\ref{G20}) for the external variable $x$ and performing
 Gauss-Chebyshev integration over the internal variables $s$ and $t$,
  (\ref{G47}) takes the form of a matrix equation
 \begin{equation}
         \begin{split}
 \left\{\bm{1}+\Quarter (b-a)^2\,\bm{W}^-\cdot\bm{W}^-\circ [p]\right\}\cdot[y]=
 \hspace*{2cm}\\
         = y(a)[1]+y'(a)[x-a] 
         + \Quarter (b-a)^2\,\bm{W}^-\cdot \bm{W}^- \cdot [q] 
         \end{split}
         \label{G48}
 \end{equation}
 We use the notation where a vector $[f]$ contains the values of
 the function $f(x)$ at 
 the mesh points (\ref{G20}), i.e.   $[f]$ 
 stands for the array $[f(x_1),f(x_2),\dots,f(x_N)]^\dagger$. The symbol 
 and $[\bm{A}\circ\bm{B}]_{ij}\equiv A_{ij}B_{ij}$
 denotes the Schur product of the matrices $\bm{A}$ and $\bm{B}$
 and $\bm{1}$ is a unit matrix.
 Writing  the Schr\"{o}dinger equation (\ref{G46}) in the form (\ref{G48}), 
 the problem has been reduced to that of solving a linear system easily
 handled by standard methods. When the vector $[y]$ has been obtained
 by solving the system (\ref{G48}), the function $y(x)$ can be reconstructed
 at any point $x\in[a,b]$ with the aid of the interpolative formula (\ref{G20a}).
Knowing $y(x)$, the derivative $y^\prime(x)$ can be
calculated either by Chebyshev differentiation (\ref{GA3})
or from the integral expression   
\begin{equation}
y^\prime(x)=y^\prime(a) - \int_a^x\D t\, p(t)\,y(t)  
         +\displaystyle\int_a^x \D t\, q(t)
   \label{G48n}
\end{equation} 
 which after discretization involves only a matrix times vector multiplication.
 The derivative is then obtained by interpolation.  
It has to be noted that in the above considerations the interval $[a,b]$ has been arbitrary. This
means that eq. (\ref{G46}) may be solved using composite Chebyshev integration that 
introduces a partition of the interval $[a,b]$. Performing integration in the first subinterval
provides us with the values of $y(x)$ and $y'(x)$ at the end-point. 
 Subsequently, they can be used to specify
 the initial conditions for the integration in the second subinterval, and so on until
 integration in the last partition has been accomplished.  This option might be particularly
useful when the functions $p(x$) and $q(x)$ are rapidly varying and their
representation would
require exceedingly large  Chebyshev mesh.  Composite integration is
 in such a case the simplest remedy.
\par     
With different boundary conditions the procedure is similar. Often we know $y(a)$ and $y^\prime(b)$
in which case eq. (\ref{G46}) is first integrated from $x$ to $b$ and next from $a$ to $x$. 
As a result,  we obtain
  \begin{equation}
         y(x)-y(a) - y'(b)(x-a)-\int_a^x\D s\displaystyle\int_s^b \D t\, p(t)\,y(t)  
         =-\displaystyle\int_a^x\D s\displaystyle\int_s^b \D t\, q(t).  
         \label{G47a}
 \end{equation} 
 Putting the external variable $x$ on the  Chebyshev mesh (\ref{G20})
 and  using Gauss-Chebyshev quadratures, (\ref{G47a}) takes the form 
 \begin{equation}
         \begin{split}
 \left\{\bm{1}-\Quarter(b-a)^2\,\bm{W}^-\cdot\bm{W}^+\circ [p]\right\}\cdot[y]=
 \hspace*{2cm}\\
         = y(a)[1]+y'(b)[x-a] 
         - \Quarter(b-a)^2\,\bm{W}^-\cdot\bm{W}^+ \cdot [q]. 
         \end{split}
         \label{G48a}
 \end{equation}
 It is evident that
 also in this case we end up with a linear system which, naturally, 
 is different from (\ref{G48}). In either case, when the vector $[y]$ has been
 determined the ultimate solution of (\ref{G46}) is given by the
 interpolative formula (\ref{G20a}).
\subsection{Volterra integral equation of the second kind}
We write the Volterra integral equation of the second kind, as
\begin{equation}
        y(x)-\lambda\int_a^x k(x,s)\,y(s)\,\D s = f(x)
        \label{G23}
\end{equation}
where $a\leq s,x \leq b$ and $\lambda$ is a given parameter,
Both, the function $f(x)$ and the kernel
$k(x,s)$ are also assumed to be provided, whereas  
 the function $y(x)$ is the sought for solution of (\ref{G23}). 
Using Gauss-Chebyshev quadrature in (\ref{G23}), 
we are led to a system of $N$ linear algebraic equations
\begin{equation}
        \left\{\bm{1}-\lambda\Half(b-a)\,
        \bm{K} \circ \bm{W}^- \right\}\cdot[y]=[f].   
        \label{G24}
\end{equation}
The solution $[y]$ of the system (\ref{G24}) used in the interpolative formula
(\ref{G20a}) yields $y(x)$ at an arbitrary point from the $[a,b]$ interval.
\par
When the independent variable $x$ appears instead as the lower integration limit, i.e. 
if (\ref{G23}) is replaced by the Volterra equation
\begin{equation}
        y(x)-\lambda\int_x^b k(x,s)\,y(s)\,\D s = f(x)
        \label{G23a}
\end{equation}
the procedure is similar and the only change required is $\bm{W}^-\to\bm{W}^+$. 
In the end we arrive at the system of equations 
\begin{equation}
        \left\{\bm{1}-\lambda\Half(b-a)\,
        \bm{K} \circ \bm{W}^+ \right\}\cdot[y]=[f].  
        \label{G24a}
\end{equation}
\subsection{Fredholm integral equation of the second kind}
The Fredholm integral equation of the second kind can be written, as
\begin{equation}
        y(x)-\lambda\int_a^b k(x,s)\,y(s)\,\D s = f(x)
        \label{G28}
\end{equation}
where $f(x)$ and $k(x,s)$ are known functions assumed to be regular in $[a,b]$
whereas $y(x)$ is the unknown function.
 The Chebyshev method is formally  akin to the Nystrom method
in which Gauss-Chebyshev quadrature has been adopted.
Similarly as in the Volterra eq. case just considered, we end up with a system of linear equations 
\begin{equation}
        \left\{\bm{1}-\lambda\Half(b-a)\,
        \bm{K} \circ w \right\}[y]=[f].
        \label{G29}
\end{equation}
 where $ [\bm{K} \circ w]_{ij}=k(x_i,x_j)\,w_j $
 with the weights $w_j$ given by (\ref{G15}). 
 \par
In many physical applications the kernel $k(x,s)$ in (\ref{G28}) might be continuous on the
diagonal $x=s$ but the derivative of the kernel
exhibits a discontinuity at this point. A typical situation occurs when
\begin{equation}
        k(x,s)=
        \begin{cases}
                k_1(x,s)\quad & \text{for $s\leq x$}\\
                k_2(x,s)\quad & \text{for $s \geq x$} 
        \end{cases}     
        \label{G26}     
\end{equation}
and if $k_1$ and $k_2$ are different functions, the kernel is referred to
as semi-continuous. The Chebyshev method is well suited to handle such
situation and the appropriate extension of (\ref{G29}), reads
\begin{equation}
        \left\{\bm{1}-\lambda\,\Half (b-a)
        \,\left( \bm{K}^{(1)} \circ \bm{W}^-
        + \bm{K}^{(2)} \circ \bm{W}^+\right) \right\}\cdot[y]=[f]    
        \label{G27}
\end{equation}
where $\bm{K}^{(1)}_{ij}=k_1(x_i, x_j)$ and $\bm{K}^{(2)}_{ij}=k_2(x_i, x_j)$. 
\subsection{Singular integral equation}
The singular integral equation 
with a Cauchy type singularity, can be written as
\begin{equation}
        y(x)-\lambda\int_a^b \dfrac{k(x,s)}{s-z}\,y(s)\,\D s = f(x)
        \label{sing7}
\end{equation}
where $z$ is a real parameter and
the integral is defined in the principal value sense.   
When $z$ falls outside the integration limits the
integral is not singular and eq. (\ref{sing7})  becomes a Fredholm equation 
considered above. Therefore, we confine our attention to the case when
$z$ lies within the 
integration limits  ($b>z>a$).
Apart from the pole at
$s=z$ the kernel
is assumed to be regular on the real axis in the integration range.
The singular quadrature rule introduced in (\ref{sing1}) allows to 
apply the Chebyshev method in the same way as in the Fredholm equation case.
This procedure may be viewed as an extension of the Nystrom approach and
 we end up with a system of linear algebraic equations 
\begin{equation}
        \left\{\bm{1}-\lambda\,
                \bm{K} \circ \omega(\tau) \right\}\cdot[y]=[f].
        \label{sing8}
\end{equation}
where $ [\bm{K} \circ \omega(\tau)]_{ij}=k(x_i,x_j)\,\omega_j (\tau)$ and
$\tau = [2z-(b+a)]/(b-a)$.
Formally, the solution of eq. (\ref{sing7}) and the solution of (\ref{G28})
look very much the same and what differs them are the weighting factors:
 the singular quadrature employs dedicated $z$-dependent weights
 $\omega_j(\tau)$ given in (\ref{sing2}). For a singular integral equation,
 however, the Fredholm alternative does not hold and the particular solution of the
 inhomogeneous equation obtained above
 has to be supplemented by the general solution of
 the homogeneous equation which must be obtained by a different method.
 \par
 The weakly singular integral equations can be handled in a similar manner. 
 We shall mention briefly the case when the kernel has a logarithmic singularity 
\begin{equation}
        y(x)-\lambda\int_a^b k(x,s)\,\log|s-z|\,y(s)\,\D s = f(x)
        \label{sing7a}
\end{equation}
where $a\leq z\leq b$ and  the function $k(x,s)$ is taken to be
regular within the integration range. The integral in (\ref{sing7a}) can be
approximated by the automatic quadrature (\ref{sing13}) and we are led
to a linear system of algebraic equations 
\begin{equation}
        \left\{\bm{1}-\lambda\Half(b-a)\,
                \bm{K} \circ [w\,\log\Half(b-a) + \Omega(\tau)] \right\}\cdot[y]=[f].
        \label{sing8a}
\end{equation}
where $ \{\bm{K} \circ [w\,\log\Half(b-a)+ \Omega(\tau)]\}_{ij}=
k(x_i,x_j)\,[w_j\,\log\Half(b-a)+\Omega_j(\tau)]$. Here $w_j$ are the Gauss-Chebyshev weights 
 and the dedicated
 weights $\Omega_j(\tau)$ are given in (\ref{sing14}) with 
$\tau = [2z-(b+a)]/(b-a)$. 
 \par
 An example of (\ref{sing7}) is the Omnes equation \cite{omnes} which arises in the
 dispersion theory of final state interaction. The equation (\ref{sing7a}) appears
 in the Faddeev approach to a three-body problem \cite{liu}. Although the above
 results could be immediately applied in these problems,
 a discussion of these
 topics is beyond the scope of the present paper.
\subsection{Integro-differential equation}
The Chebyshev method is also very well suited to handle 
integro-differential equations. As an example we may consider the
equation
 \begin{equation}
         y'(x)+p(x)\,y(x)=q(x)+\int_a^b k(x,s)\,y(s)\,\D s
         ;\quad a\leq x\leq b
         \label{G49}
 \end{equation}
 where $p(x), q(x)$ and $k(x,s)$ are given functions. We also assume
 that the value of $y(a)$ has been assigned.
 Upon integration of (\ref{G49}) in the limits from $a$ to $x$, we obtain
 \begin{equation}
         y(x)-y(a) + \int_a^x p(s)\,y(s) \,\D s 
         =\displaystyle\int_a^x q(s)\,\D s+ 
         \displaystyle\int_a^x\displaystyle\int_a^b k(t,s)\,y(s)\,\D t\,\D s.
         \label{G50}
 \end{equation}
 Introducing  Chebyshev mesh, we are led to a linear system
 \begin{equation}
         \left\{\bm{1}+
     \Half(b-a) \bm{W}^-\circ [p] - \Half(b-a)\bm{W}^-\cdot\bm{K}\circ [w]
     \right\}
     \cdot[y] 
     = y(a)[1] 
         + \Half(b-a)\bm{W}^-\cdot [q]. 
         \label{G51}
 \end{equation}
Another type of integro-differential equation that might be encountered
in physical applications is the Schr\"{o}dniger equation with exchange interaction.  
Consider then a second order homogeneous equation
 \begin{equation}
         y''(x)+p(x)\,y(x)=\int_a^b k(x,s)\,y(s)\,\D s
         ;\quad a\leq x\leq b
         \label{G52}
 \end{equation}
 where $p(x), q(x)$ and $k(x,s)$ are given functions. We also assume
 that $y(a)$ and $y'(a)$ take assigned values. 
 When (\ref{G52}) is integrated twice from $a$ to $x$, we obtain
 \begin{equation}
         y(x)-y(a) - y'(a)(x-a)+\int_a^x\D s\displaystyle\int_a^s \D t\, p(t)\,y(t)  
        = \displaystyle\int_a^x\displaystyle\int_a^b k(t,s)\,y(s)\,\D t\,\D s.
         \label{G53}
 \end{equation}
 Upon introducing the Chebyshev mesh, we end up with a liner system of equations
 \begin{equation}
         \begin{split}
         \left\{\bm{1}-\Quarter(b-a)^2\,\bm{W}^-
         \cdot\bm{W}^-\circ [p] -
         \Quarter(b-a)^2\,\bm{W}^-\cdot\bm{K}\circ [w]\right\}\cdot[y]
         =\hspace*{1cm}\\ 
         = y(a)[1]+y'(a)[x-a]. 
 \end{split}
         \label{G54}
 \end{equation}
 In the Chebyshev approach the inclusion of exchange forces brings
 only a minor complication. The amount of labor required to obtain the   
 solution of the integro-differential equation (\ref{G52})
 is about the same as in the case
 of the differential equation (\ref{G46}).
 \section{ Quantum mechanical two-body problem }          
 \subsection{Configuration space}
 We wish to examine the effectiveness of 
 the pseudo-spectral Chebyshev method by
 solving explicitly a number of scattering and 
 bound states problems  
 but before we do that, we need to establish our notation 
 and assemble some well known theoretical tools.  
 For the sake of clarity, we confine our attention to a single
 channel, time independent two-body problem. To simplify matters even further,
 we assume that the ''particles'' are deprived of any internal
 degrees of freedom.  The adopted interaction has a form of a   
 local, short-ranged, spherically symmetric 
 potential $V(r)$ which depends only
 upon the mutual separation $r$ between the particles. 
 The shape of $V(r)$ is arbitrary and a singular behavior
 when $r\to 0$  is not excluded with the  
 restriction that this singularity would be not worse than $1/r$. 
 Since the $1/r$ singularity requires special care, 
 at the top of  $V(r)$ we introduce explicitly a
 point charge Coulomb potential
 $V_C(r)=\alpha Z/r$ where $\alpha$ is the fine structure constant and $Z$ is
 the charge number. 
 \par
 The radial part of the appropriate Schr\"{o}dinger equation, is 
 \begin{equation}        
 \D^2 \psi(p,r)/\D r^2+\left[\,p^2
 -\ell(\ell+1)/r^2 -2\mu\,V_C(r) \right]\psi(p,r)
 = 2\mu\,V(r)\, \psi(p,r),
         \label{QM1}
 \end{equation}
 where $p,\;\mu$ and $\ell$ denote, respectively,
 the center-of-mass momentum,  the reduced mass and 
 the orbital momentum. Since in (\ref{QM1}) different $\ell$ do not mix, 
 the label $\ell$ on the wave function $\psi(p,r)$ has been dropped.
 The regular solution of (\ref{QM1}) is defined by 
 the boundary condition 
 \begin{equation}        
         \psi(p,r)\sim r^{\ell+1} \quad \text{for}\quad r\to 0.
         \label{QM2}
 \end{equation}
 In a scattering problem the momentum $p$ takes real non-negative values
 whereas in a bound state problem the momentum will be imaginary. 
 We shall deal with both problems in sequence
 commencing with the continuous spectrum case where our objective is
 to calculate the scattering phase shift.
 \par
 In absence of $V(r)$, the two linearly independent 
 solutions of (\ref{QM1}), and referred to 
 in the following as the ''free solutions'',  are
 \begin{eqnarray*}
 && f(\rho)\equiv F_\ell(\eta,\rho),\\  
 && g(\rho)\equiv G_\ell(\eta,\rho), 
 \end{eqnarray*}
 where $F_\ell(\eta,\rho)$ and $G_\ell(\eta,\rho)$ 
 are the standard Coulomb wave functions defined in \cite{abramowitz}, $\rho=p\,r$ and
 $\eta=\mu\alpha Z/p$ is the familiar Sommerfeld parameter.
 When both, the Coulomb interaction is switched off ($Z=0$), 
 and $V(r)=0$, the free solutions simplify to the form 
 \begin{eqnarray*}
 && f(\rho)\equiv\rho\,j_\ell(\rho),\\  
 && g(\rho)\equiv-\rho\,n_\ell(\rho), 
 \end{eqnarray*}
 where $j_\ell$ and $n_\ell$
 denote the usual spherical Bessel functions.
 \par
 For $r>R$ where $R$ 
 is some cutoff radius which is sufficiently large as compared
 with the range of the potential $V(r)$, 
 the potential term on the right hand side in (\ref{QM1}) gives negligible
 contribution and $\psi(p,r)$ is a superposition of the free solutions
 \begin{equation}
        \psi(p,r)\approx A[f(pr)+\tan\delta\,g(pr)],\quad \text{for}\quad r>R,
         \label{QM6}
 \end{equation}
 where $A$ is a constant amplitude and $\delta$ denotes the phase shift  
 (suppressing the $\ell$ label). Given the logarithmic derivative of
 $\psi(p,R)$, the phase shift is computed from
 \begin{equation}
         \tan\delta=-\dfrac{f(pR)\,\psi^\prime(p,R)/\psi(p,R)-f^\prime(pR)}
                           {g(pR)\,\psi^\prime(p,R)/\psi(p,R)-g^\prime(pR)}
         \label{QM7}
 \end{equation}
 where prime denotes the derivative with respect to $R$.
 The free solutions can be also
 utilized in constructing the standing-wave Green's function $G_p(r,r')$
 \begin{equation}
         G_p(r,r')=-(1/p)\,f(pr_{\scriptscriptstyle<})\,g(pr_{\scriptscriptstyle>}) 
         \label{QM3}
 \end{equation}
 where
         $r_{\scriptscriptstyle<}=\min(r,r')$ and
         $r_{\scriptscriptstyle>}=\max(r,r')$.
 Because the Green's function (\ref{QM3}) is a solution of the
 inhomogeneous wave equation
 \begin{equation}        
 \D^2 G_p(r,r')/\D r^2+\left[\,p^2
 -\ell(\ell+1)/r^2 -2\mu\,V_C(r) \right]\,G_p(r,r')
 = \delta(r-r'),
         \label{QM4}
 \end{equation}
 the Schr\"{o}dinger equation (\ref{QM1}) may be
 immediately converted into the L-S equation
 \begin{equation}        
        \psi(p,r) = f(pr) +  
                2\mu\,\int_0^\infty G_p(r,r')\,V(r')\,\psi(p,r')\,\D r'.
         \label{QM5}
 \end{equation}
 This integral equation
 incorporates the boundary conditions (\ref{QM2}) and (\ref{QM6}).
 Since the incident wave has amplitude equal to one, 
 given the solution of (\ref{QM5}), the phase shift is obtained from the formula
 \begin{equation}
        \tan\delta= -\frac{2\mu}{p}\int_0^\infty f(pr)\,V(r)\,\psi(p,r)\,\D r.
        \label{QM8}
 \end{equation}
 It will be advantageous
 casting the  equation
 into a Volterra equation. Indeed, the Fredholm
 equation (\ref{QM5}) can be rewritten, as
\begin{equation}
        \psi(p,r)= C\,f(pr) 
         -\frac{2\mu}{p} \int_0^r \left[ f(pr')\, g(pr) - g(pr')\, f(pr)\right]
        \,V(r')\,\psi(p,r')\,\D r' 
        \label{QM9}
\end{equation}
 where
\begin{equation}
        C \equiv 1- \frac{2\mu}{p} \int_0^\infty g(pr)\,V(r)\,\psi(p,r)\,\D r
        \label{QM10}
\end{equation}
 is a constant which does not depend on $r$.  
 At first sight, (\ref{QM9}) involving an unknown constant $C$ 
 does not appear to be of much use, but in fact it is possible  
 to get rid of $C$  and eventually pin down its value. 
 Indeed, introducing a new wave function function $u(p,r)$, 
 defined by the formula $\psi(p,r)=C\,u(p,r)$,  eq.  
 (\ref{QM9}) reduces to a Volterra equation of the second kind for the function $u(p,r)$
 and this equation is free from unknown constants
\begin{equation}
        u(p,r)=f(pr) 
        -\frac{2\mu}{p} \int_0^r \left[ f(pr') \,g(pr) - g(pr') \, f(pr)\right]
        \,V(r')\,u(p,r')\,\D r'. 
        \label{QM11}
 \end{equation}
 Solving  eq. (\ref{QM11}) for $u(p,r)$, 
 the value of $C$ can be determined  
 by first substituting $\psi=Cu$ in (\ref{QM10}) and
 then  solving the resulting equation for $C$. We get
 \begin{equation}
        C = \left\{1 + \frac{2\mu}{p} \int_0^\infty g(pr)\,V(r)\,u(p,r)\,\D r\right\}^{-1}
        \label{QM12}
 \end{equation}
 and the ultimate formula for the phase shift 
 obtained by substituting $\psi=Cu$ in  (\ref{QM8}), is
 \begin{equation}
  \tan\delta= -C\;\dfrac{2\mu}{p}\displaystyle\int_0^\infty f(pr)\,V(r)\,u(p,r)\,\D r,
        \label{QM13}
 \end{equation}
 with $C$ given in (\ref{QM12}).
 This formula makes reference only to the solution of eq. (\ref{QM11}).
 The possibility that the configuration space non-relativistic scattering problem 
 might be formulated in terms of an inhomogeneous
 Volterra equation of the second kind was first noted by Drukarev \cite{drukarev}
 more than half a century ago. 
 Although his method soon received the necessary mathematical background \cite{brysk}, 
 the Volterra equation approach went into oblivion to be revived only recently
 \cite{gonzales}\cite{kang1}\cite{kang2}.
 \par
 The above scheme based on the Volterra integral equation may be also applied
 for calculating the scattering length but
 the calculation of the Coulomb corrected scattering length is slightly more complicated
 and for this purpose one needs a different set of Coulomb wave functions which are analytic
 and entire functions of $p^2$. Such functions, denoted here as
 $\Phi_\ell(p, r)$ and $\Theta_\ell(p,r)$, are
  particular superpositions
 of $F_\ell$ and $G_\ell$, and are available in the literature. 
 Following \cite{lambert}, for $p=0$ the regular function takes the form
 \begin{equation}
 \label{e13}
 \Phi_{\ell}(0,r)=
 \begin{cases}
 [(2\ell+1)!/\beta^{\ell+1}]\sqrt{\beta\,r}
                  \;   I_{2\ell+1}(2\sqrt{\beta\,r})  &  Z>0\\
  r^{\ell+1}                                          &  Z=0\\
  [(2\ell+1)!/\beta^{\ell+1}]\sqrt{\beta\,r}
                  \;   J_{2\ell+1}(2\sqrt{\beta\,r})  &  Z<0
 \end{cases} 
 \end{equation}
 where $\beta=2\mu \alpha |Z|$, and $(J_{2\ell+1},I_{2\ell+1})$ are
 Bessel and modified Bessel functions, respectively \cite{abramowitz}.
 The irregular function $\Theta_{\ell}(0,r)$
 is given in terms of the Neuman $Y_{2\ell+1}$ and modified
 Neuman function $K_{2\ell+1}$, respectively \cite{abramowitz}
 \begin{equation}
 \label{e15}
 \Theta_{\ell}(0,r)=
 \begin{cases}
  2\,[\beta^\ell/(2\ell+1)!]\sqrt{\beta\,r}
  \;  K_{2\ell+1}(2\sqrt{\beta\,r})  &    Z>0  \\
  r^{-\ell}/(2\ell+1)  &    Z=0 \\
 -\pi\,[\beta^{\ell}/(2\ell+1)!]\sqrt{\beta\,r}
  \;  Y_{2\ell+1}(2\sqrt{\beta\,r})  &   Z<0
 \end{cases}
 \end{equation}
 and with the adopted normalization the Wronskian $W[\Theta_\ell,\Phi_\ell]$ is
 equal unity.
 The Coulomb corrected scattering length $A_C$ is then obtained
 \cite{lambert} from the formula
 \begin{equation}
         A_C= 
 -2\mu\displaystyle\int_0^\infty \Phi_0(0,r)\,V(r)\,\phi(r)\,\D r
  \left\{1 + 2\mu\displaystyle \int_0^\infty \Theta_0(0,r)\, V(r)\,\phi(r)\,\D r\right\}^{-1},
        \label{QM13ac}
 \end{equation}
 where the function $\phi(r)$ is a solution of the Volterra equation 
\begin{equation}
        \phi(r)= \Phi_0(0,r) 
        -2\mu \int_0^r [\Phi_0(0,r')\,\Theta_0(0,r)-\Phi_0(0,r)\,\Theta_0(0,r')]
        \,V(r')\,\phi(r')\,\D r'. 
        \label{QM11ac}
 \end{equation}
 The above scheme remains valid in absence of Coulomb interaction.
 \par 
Although the method outlined above does not seem to have been tried for bound state
calculations, such extension is perfectly feasible.  Bound states are identified with poles
of the T-matrix that are located on the positive part of the imaginary axis
in the complex momentum plane. Therefore, 
formula (\ref{QM13}) must be replaced by the appropriate T-matrix expression
 in which the physical momentum $p$ will be analytically
continued to the complex momentum plane $k$. 
Using the T-matrix scheme for scattering 
requires the outgoing wave Green's function
$ G^{(+)}_p(r,r')=-
 h(pr_{\scriptscriptstyle>}) \;f(pr_{\scriptscriptstyle<})/p$ 
 where $h(x)=g(x)+\I\,f(x)$.
 The corresponding  wave function $\psi^+(p,r)$ is obtained 
as the solution of the  equation
\begin{equation}
        \psi^+(p,r)=f(kr)+2\mu\,\int_0^\infty G^{(+)}_p(r,r')\,V(r')\,\psi^+(p,r')\,\D r'.
        \label{B1}
\end{equation}
Passing to the asymptotic limit $r\to\infty$ in (\ref{B1}) 
immediately yields the on-shell T-matrix formula
\begin{equation}
        T(p)= \E^{\I\delta} \,\sin\delta= 
        -\dfrac{2\mu}{p}\displaystyle\int_0^\infty f(pr)\,V(r)\,\psi^+(p,r)\,\D r.
        \label{B2}
\end{equation}
Similarly as before, eq. (\ref{B1}) can be converted to a Volterra equation 
\begin{equation}
        \psi^+(p,r)= D\,f(pr) 
         -\frac{2\mu}{p} \int_0^r 
                 \left[ f(pr') g(pr) - g(pr') f(pr)\right]
        \,V(r')\,\psi^+(p,r')\,\D r' 
        \label{B3}
\end{equation}
 which has the same kernel as (\ref{QM9}) but
 the constant $D$ is different from $C$ and reads    
\begin{equation}
        D = 1 - \frac{2\mu}{p} \int_0^\infty h(pr)
                \,V(r)\,\psi^+(p,r)\,\D r.
        \label{B4}
\end{equation}
The unknown constant $D$ may be removed by setting $\psi^+(p,r)=D\,u(p,r)$ 
where the function $u(p,r)$ is a solution of (\ref{QM11}), the very same equation
which was used in the calculation of $\tan\delta$.
Eliminating $\psi^+$ in favor of $u$ in (\ref{B2}), 
 the ultimate T-matrix formula takes the form 
\begin{equation}
       T(p) = 
    -\dfrac{2\mu}{p}\displaystyle\int_0^\infty f(pr)\,V(r)\,u(p,r)\,\D r
   \left\{1 + \dfrac{2\mu}{p}\displaystyle
   \int_0^\infty h(pr) \,V(r)\, u(p,r)\,\D r\right\}^{-1}.
        \label{B5}
\end{equation}
 The T-matrix (\ref{B5})  
 satisfies elastic unitarity constraint $\Im\,T^{-1}=-1$.
 The expression occurring in the denominator of (\ref{B5}) is recognized as the Fredholm
 determinant whose zeros located on the positive imaginary axis in the complex k-plane 
 would be identified with bound states.
 Therefore, we need to make an analytic
 continuation from the physical region onto the imaginary axis 
 $ p\to\I\kappa$ where $\kappa$ is real and non-negative.
 As a result, the Coulomb wave functions 
 $F_\ell$ and $H^+_\ell$ will be modified, and we have
 \begin{eqnarray}
  F_\ell(\eta,pr)  &\to& \I^{\ell+1}\; \exp{( \I\,\Half\pi \tilde{\eta})}\; \tilde{f}(\kappa r),
  \label{B5e}\\
  H^+_\ell(\eta,pr)&\to& \I^{-\ell }\; \exp{(-\I\,\Half\pi \tilde{\eta})}\; \tilde{h}(\kappa r),
  \label{B5f}
 \end{eqnarray}
 where $\tilde{\eta}=\mu \alpha Z/\kappa$. The new functions $\tilde{f}$ and $\tilde{h}$
 are both real and are given in analytic form
  \begin{eqnarray}
            \tilde{f}(\kappa r)&=&C\, 
          (2\kappa r)^{\ell+1}\;\E^{-\kappa r} \; M(\ell+1+\tilde{\eta},\,2\ell+2,\,2\kappa r);
          \label{B5c}     \\
        \tilde{h}(\kappa r)&=&
          (2\kappa r)^{\ell+1}\;\E^{-\kappa r} \; U(\ell+1+\tilde{\eta},\,2\ell+2,\,2\kappa r),
           \label{B5d}   
  \end{eqnarray}
  with $C=|\Gamma(\ell+1+\tilde{\eta})|/[2(2\ell+1)!]$ where $M(a,b,x)$ and $U(a,b,x)$ denote,
  respectively, the Kummer and Tricomi confluent hypergeometric functions 
  \cite{abramowitz} in which
  all arguments are real
  (numerical methods of calculating negative energy Coulomb wave functions
  have been presented in \cite{noble1},\cite{noble2} and \cite{koval}).
  In the charge-less case ($Z=0$), the above functions simplify to the form
 \begin{eqnarray}
        && \tilde{f}(\kappa r)  =  (\kappa r)\,i_\ell(\kappa r);
        \label{B5a}\\
    &&\tilde{h}(\kappa r)  = (2/\pi)(\kappa r)\,k_\ell(\kappa r),
        \label{B5b}
\end{eqnarray}
where $i_\ell(x)$ and $k_\ell(x)$ are the modified spherical Bessel functions 
defined in \cite{abramowitz}.  The adopted normalization is such that
the Wronskian $\tilde{h}(x)\,\tilde{f}'(x)-\tilde{f}(x)\,\tilde{h}'(x)$ is equal 
to unity.
\par
The kernel in (\ref{QM9}) analytically continued onto the
imaginary axis remains real and 
we infer that $u(\I\kappa,r)$ acquires merely a constant phase factor equal 
 $  \I^{\ell+1}\, \exp{ (\I\,\Half\pi \tilde{\eta}) } $.
 Therefore,  we set
 $u(\I\kappa,r)=
 \I^{\ell+1}\, \exp{( \I\,\Half\pi\tilde{\eta})} \,\tilde{u}(\kappa,r)$ 
 with real $\tilde{u}$.  On the imaginary axis
 the Fredholm determinant $\Delta(\kappa)$ is real, and we have
\begin{equation}
      \Delta(\kappa)=
          1 + \dfrac{2\mu}{\kappa}\; 
                  \int_0^\infty \tilde{h}(\kappa r)\,V(r)\,
                \tilde{u}(\kappa,r)\,\D r,
        \label{B6}
\end{equation}
where $\tilde{u}$ is the solution of the Volterra equation involving only real
quantities
\begin{equation}
        \tilde{u}(\kappa,r)= \tilde{f}(\kappa r) 
         -\dfrac{2\mu}{\kappa} \int_0^r 
                 [\tilde{f}(\kappa r') \,\tilde{h}(\kappa r) 
                 - \tilde{h}(\kappa r') \,\tilde{f}(\kappa r)]\,V(r')
   \,\tilde{u}(\kappa,r')\,\D r'.
        \label{B7}
\end{equation}
 The bound states are located by finding the roots of the equation
 \begin{equation}
      \Delta(\kappa)=0.
         \label{B8}
 \end{equation}
 Denoting such root as $\kappa_0$ the corresponding binding
 energy $B$ will be  obtained as $B=-\kappa_0^2/2\mu$. 
 \par
 In the literature, the usual method of locating the binding energies
 utilizes the homogeneous  equation (\ref{B1}). The configuration space
 is rather unwieldy for this purpose because the Green's function
 exhibits a cusp behavior at $r=r'$, which in general
 leads to a loss of accuracy.
 Therefore, momentum space methods have been given preference and 
 the task of solving the homogeneous
  equation reduces to that of an  
 algebraic eigenvalue problem easily handled by standard methods.
 Although, the pseudo-spectral method can cope with such
 a situation when the kernel is semi-continuous (\ref{G27}),
 but owing to the presence of the
 $\bm{W}^\pm$ matrices the resulting eigenvalue problem 
 in configuration space would involve non-symmetric matrices. 
 Unfortunately, the standard methods of solving eigenvalue problems are
 in this case much less accurate than for symmetric matrices, rendering the 
 configuration space method a less attractive option. Thus, in the homogeneous
  equation case the momentum space approach is still preferable.
 \subsection{Momentum space}
 In momentum space the  partial wave Schr\"{o}dinger equation, takes the form
 \begin{displaymath}
         k^2\;\phi(k)+\dfrac{2}{\pi}\int_0^\infty\,U_\ell(k,k')\,\phi(k')\,k^{\prime 2}\,\D k'
        =2\mu\,E\,\phi(k),
 \end{displaymath}
 where $U_\ell(k,k')$ denotes the
 $\ell$-th wave projection of the local potential $2\mu \, V(r)$ 
  \begin{equation}
          U_\ell(k',k)=2\mu \int_0^\infty j_\ell(k'r)\,V(r)\,j_\ell(kr)\,r^2\,\D r.
          \label{sing10}
  \end{equation}
 Setting $E=-\kappa^2/2\mu$ for a bound state problem and introducing a new unknown function 
 \begin{displaymath}
         u(k)=k\,\sqrt{k^2+\kappa^2}\,\phi(k)
 \end{displaymath}
 we obtain a homogeneous Fredholm integral equation  
 \begin{displaymath}
         u(k)=\int_0^\infty {\cal K}(k,k')\,u(k')\,\D k',
 \end{displaymath}
 with a manifestly symmetric kernel
 \begin{displaymath}
         {\cal K}(k,k') = -(2/\pi)\,(k/\sqrt{k^2+\kappa^2})\,U_\ell(k,k')\,
         (k'/\sqrt{k^{\prime 2} +\kappa^2}).
 \end{displaymath}
 Following the customary procedure, 
 this kernel may be multiplied by a constant ''eigenvalue''  $\lambda$
 which for an assigned $\kappa$  is in fact a function $\lambda(\kappa)$.
 At a particular value $\kappa=\kappa_0$ for which $\lambda(\kappa_0)=1$
 we would have a bound state. In practice, after discretization we end up with
 an algebraic eigenvalue problem in which $\kappa$ is a parameter. 
   \par
  The solution to the non-relativistic potential theory scattering
  problem is obtained
  from the L-S equation for the t-matrix. It is much easier 
  to deal with real quantities
  and therefore a convenient starting point is the partial wave L-S equation
  for the off-shell K-matrix
  \begin{equation}
  \bra{k'}K_\ell(p^2)\ket{k}=U_\ell(k',k)-\dfrac{2}{\pi}\int_0^\infty     
  U_\ell(k',q) \; \dfrac{q^2\,\D q}{q^2-p^2}\; \bra{q}K_\ell(p^2)\ket{k}
          \label{sing9}
  \end{equation}
  which is valid for spherically symmetric potentials. 
  When eq. (\ref{sing9}) has been solved, the 
   t-matrix can be recovered from
  the off-shell unitarity constraint and 
  the resulting fully off-shell t-matrix reads
  \begin{equation}
  \bra{k'}T_\ell(p^2\pm \I\epsilon)\ket{k}      = 
  \bra{k'}K_\ell(p^2)\ket{k} \mp \I p \; 
  \dfrac{         
  \bra{k'}K_\ell(p^2)\ket{p}\; \bra{p}K_\ell(p^2)\ket{k} }        
  {1\pm \I p\;\bra{p}K_\ell(p^2)\ket{p} }.        
          \label{sing11}
  \end{equation}
  Finally,
  the on-shell K-matrix is related to the phase shift $\delta_\ell(p)$, by
  \begin{equation}
	  \bra{p}K_\ell(p^2)\ket{p} = -p^{-1}\,\tan\delta_\ell(p).          
          \label{sing12}
  \end{equation}
  The presence of the Coulomb potential creates the well known conceptual difficulties,
  and e.g., the Coulomb scattering amplitude becomes singular on-shell. 
  For  $(V+V_C)$, a two-potential problem the scattering amplitude can be written
  rigorously as a sum of the purely Coulomb amplitude plus the Coulomb corrected 
  amplitude associated with the short-ranged potential.  
  To escape the difficulties inherent in  the {\it ab initio}
  calculation of the Coulomb amplitude in momentum space, the latter is usually
  regarded as known.  When this is accepted, the remaining
  amplitude can be obtained quite easily from
  an extended L-S equation in which the Coulomb Green's function replaces the free Green's
  function.  For a repulsive Coulomb interaction  
  the spectral representation of the Coulomb Green's function 
  contains only the continuous spectrum and 
  the L-S equation for the Coulomb corrected 
  T-matrix retains its standard form 
  \begin{equation*}
          \bra{k'}T_\ell(p^2)\ket{k}_c
          =U_\ell(k',k)_c-\dfrac{2}{\pi}\int_0^\infty     
          U_\ell(k',q)_c \; \dfrac{q^2\,\D q}{q^2-p^2-\I\epsilon}\; 
          \bra{q}T_\ell(p^2)\ket{k}_c
  \end{equation*}
  where the subscript $c$ indicates that Coulomb representation has been used. In particular,
  the potential term in this representation is given as
  \begin{equation}
          U_\ell(k',k)_c=
          (2\mu/kk') \int_0^\infty F_\ell(\eta_{k'},\, k'r)\;V(r)
          \;F_\ell(\eta_k,\, kr)\;\D r  
  \end{equation}
  and involves explicitly the Coulomb wave functions with 
  $\eta_k=\mu\alpha Z/k$ and $\eta_{k'}=  \mu\alpha Z/k' $ .
  The T-matrix on-shell yields directly the Coulomb corrected phase shift
  \begin{equation*}
	  \bra{p}T_\ell(p^2)\ket{p}_c = -p^{-1}\;\exp{(\I\delta_\ell^c(p))}\,      
           \sin\delta_\ell^c(p).    
  \end{equation*}  
\section{ Solution of the Schr\"{o}dinger equation} 
\subsection{Continuous spectrum }
In principle, the scattering problem 
based on the Schr\"{o}dinger equation
could be solved by directly applying (\ref{G48})
where the appropriate boundary conditions are
$y(a)=0$ and $y^\prime(a)=1$. However, there is a better
 alternative in which the proper threshold behavior 
 of the wave function is enforced   from the onset    by setting
$\psi(p,r) = r^{\ell+1}\phi(p,r)$ where the unknown function $\phi(p,r)$
 is a solution of the differential equation
\begin{equation}
 \D^2 \phi(p,r)/\D r^2=
 -[p^2-2\mu\,V(r)]\,\phi(p,r)-[(\ell+1)/r]\,\D\phi(p,r)/\D r.
\label{G36a}
\end{equation} 
 Following the well known procedure, the derivative of the wave function 
 with respect to $r$ would be regarded as a second function to be determined  
 putting $\chi(p,r)=\phi^\prime(p,r)$. 
 The wave equation (\ref{G36a}), takes then the matrix form
 \begin{equation}
         \binom{\phi^\prime(p,r)}{\chi^\prime(p,r)}=
         \begin{pmatrix} 
                 &0 & 1\\
  & -\left[p^2 -2\mu\, V(r)\right]
                 & -2(\ell+1)/r
         \end{pmatrix} 
         \binom{\phi(p,r)}{\chi(p,r)}.
         \label{G36b}
 \end{equation}
 The system of equations (\ref{G36b}) will be solved using the pseudo-spectral method 
by introducing the Chebyshev mesh $x_j=\Half R[1+\cos(\pi(j-\Half)/N)]$
in the interval $[0,R]$. 
In order to obtain a system of linear equations
for the $[\phi]$ and $[\chi]$ vectors, we need to impose the boundary conditions at
the origin. In view of the fact that 
the centrifugal barrier factor has been already taken care of, the wave
function $\phi$ goes to a constant at the origin. Since in the calculation of the phase shift
we need only the logarithmic derivative of $\phi$, a common normalizing factor in
$\phi$ and $\phi^\prime$ is irrelevant,  and we set
$\phi(p,0)=1$ and $\chi(p,0)=c$ where $c$ is a constant.
For potentials which are less singular than $1/r$, 
setting $c=0$ would be sufficient to neutralize the
 singular behavior of the $(\ell+1)/r$ term.
The ultimate choice, comprising the case when 
$V(r)$ exhibits a $1/r$ singularity, is
$c=\mu\,\lim_{r\to 0}rV(r)/(\ell+1)$. Integrating  (\ref{G36b}) on 
the Chebyshev mesh,
the resulting system of $2N$ algebraic equations can be written in
matrix form
\begin{equation}
         \begin{pmatrix}
                 &\bm{1} &-\Half R\,\bm{W}^-\\
                 &\Half R\, \bm{W}^-\circ [p] & \bm{1}+\Half R\, \bm{W}^-\circ [q]
         \end{pmatrix}
  \binom{[\phi]}{[\chi]} 
  =\binom{[1]}{[c]}, 
\label{G36d}
\end{equation}
where $[p]_i= \{p^2 -2\mu\, V([x]_i)\}$ and $[q]_i=2(\ell+1)/[x]_i$.   
When the system
(\ref{G36d}) has been solved, the wave function and its derivative are provided
by the interpolative formula 
 \begin{equation}
\phi(p,r)
  =\sum_{j=1}^N {\rm '\mbox{} } \ G_j(2r/R-1)\,\phi(p, x_j),
\label{G40}
\end{equation} 
 and
 \begin{equation}
\chi(p,r)
  =\sum_{j=1}^N {\rm '\mbox{} } \ G_j(2r/R-1)\,\chi(p, x_j),
\label{G41}
\end{equation} 
 respectively. Knowing the logarithmic derivative of $\phi(p,R)$, 
the phase shift is obtained from
\begin{equation}
        \tan\delta=-\dfrac{f(pR)\,[(\ell+1)/R+\chi(p,R)/\phi(p,R)]-f^\prime(pR)}
                          {g(pR)\,[(\ell+1)/R+\chi(p,R)/\phi(p,R)]-g^\prime(pR)},
\label{G36e}
\end{equation}   
where prime denotes derivative with respect to $r$.
\par
 The above scheme may be used to calculate the scattering length
 by going to the limit $p\to 0$. Dividing both sides of
 (\ref{G36e}) by $p$ and subsequently
 by letting $p$ go to zero, we obtain the scattering length ($\ell=0$)
 \begin{equation}
         A = -R\;{\cal L}(R) /[1+ {\cal L}(R)],
         \label{G36f}
 \end{equation}
 where  ${\cal L}(R)= R\chi(0,R)/\phi(0,R)$ and  
$\phi(0,R)$ and $\chi(0,R)$ are the solutions of the system (\ref{G36b}) in
 which the momentum $p$ has been set equal to zero.
\par
When the potential $V(r)$  is given as  a sum of a short ranged ''nuclear'' potential plus
a  point charge Coulomb potential, formula (\ref{G36e}) remains valid and 
the resulting phase shift becomes then Coulomb distorted nuclear phase shift.
 In this case the free solutions $f$ and $g$  must be
replaced by the appropriate Coulomb wave functions $F_\ell(\eta,\rho)$
and $G_\ell(\eta,\rho)$, respectively, and the same must be repeated for the
derivatives.   The Coulomb corrected scattering length is given by
a more complicated expression
\begin{equation}
	A=-\{R\Phi_0'(0,R) - [1+{\cal L}(R)]\Phi_0(0,R)\}/
 \{R\Theta_0'(0,R) - [1+{\cal L}(R)]\Theta_0(0,R)\}
\label{G36g}
\end{equation}
where prime denotes the derivative with respect to $r$ and  the calculated
${\cal L}(R)$ takes also account of the Coulomb potential.
\subsection{Bound states}
 In the case of a bound state the momentum is imaginary $p\to\I\kappa$ and the 
 wave function will be a function of $\kappa$ and $r$. Our goal is to
 calculate $\kappa$  which determines the binding energy.
 In order to enforce the proper threshold behavior
 of the wave function, we set $\psi(\kappa,r) = r^{\ell+1}\,\phi(\kappa,r)$ 
 where the hitherto unknown
 function $\phi(\kappa,r)$, satisfies the Schr\"{o}dinger equation
 \begin{equation}
 \D^2 \phi(\kappa,r)/\D r^2=-[2(\ell+1)/r]\; \D \phi(\kappa,r)/\D r
 + \left[\kappa^2 +2\mu\,V(r)\right]\,\phi(\kappa,r). 
         \label{G37}
 \end{equation}
 For $V(r)=0$ eq. (\ref{G37}) has two linearly independent solutions 
$(\kappa r)^{-\ell}\,i_\ell(\kappa r)$ and $(\kappa r)^{-\ell}\,k_\ell(\kappa r)$.
 The sought for function $\phi(\kappa,r)$ is the regular solution of 
  (\ref{G37}) decaying exponentially for large $r$.
 It will be convenient replacing the single second order
 equation (\ref{G37}) by a system of two,
 first order differential equations, which is accomplished
 by introducing formally a second function $\chi(\kappa,r)$ 
 and equal to the derivative of the wave
 function $\chi(\kappa,r)=\D \phi(\kappa,r)/\D r$. 
 The wave equation (\ref{G37}), takes the matrix form
 \begin{equation}
         \binom{\phi^\prime(\kappa,r)}{\chi^\prime(\kappa,r)}=
         \begin{pmatrix} 
                 &0 & 1\\
  &  \left[\kappa^2 +2\mu\, V(r)\right]
                 & -2(\ell+1)/r
         \end{pmatrix} 
         \binom{\phi(\kappa,r)}{\chi(\kappa,r)}.
         \label{G37a}
 \end{equation}
 Although the variable $r$ varies between zero 
 and infinity, for practical reasons infinity must be replaced  
 by some cutoff radius $R$ which is 
  much bigger than the range of the potential. 
  Thus,  in the following it will be assumed
 that $r$ belongs to the interval $[0,R]$. We are interested in the regular solution of
 (\ref{G37}) and in view of the fact that the threshold factor controls
 the rate of the fall off down to zero of the wave function,
 $\phi(\kappa,r)$ must go to a constant for $r\to 0$. 
 Since this constant would be absorbed in the normalization,
 we take its value to be equal to unity.
 At $r=0$ we have  $\chi(\kappa,0)=c$ where
 $c=\mu\,\lim_{r\to 0}rV(r)/(\ell+1)$. 
 \par
 Integrating eq. (\ref{G37a}) in the limits $[0,R]$ and introducing the Chebyshev grid,   
 we are presented with a system of algebraic equations 
 \begin{equation}
         \begin{pmatrix}
                 &\bm{1}&-\Half R\,\bm{W}^-\\
                 &-\Half R\, \bm{W}^-\circ [p] & \bm{1}+ \Half R\, \bm{W}^-\circ [q]
         \end{pmatrix}
  \binom{[\phi]}{[\chi]} 
  =\binom{[1]}{[c]} 
\label{G43}
\end{equation}
where $[p]_i= \kappa^2 +2\mu\,V([x]_i);\;[q]_i=2(\ell+1)/[x]_i$. 
When the system of linear algebraic equations (\ref{G43}) has been solved, the wave function and
its derivative can be  reconstructed at any point from (\ref{G40}) and (\ref{G41}), respectively
making the substitution $p\to\kappa$. 
\par
To locate a bound state we impose the requirement
that the wave function  
 falls off exponentially at large separations. 
 In the asymptotic region  $r \geq R$ 
the potential $V(r)$ becomes negligible and the wave function must be a superposition
of the free solutions
\begin{equation}
\phi(\kappa,r)\approx (\kappa r)^{-\ell}
 \,[A(\kappa)\,k_\ell(\kappa r) + B(\kappa)\,i_\ell(\kappa r)],
\label{G40a}
\end{equation}
where the two coefficients $A(\kappa)$ and $B(\kappa)$ may be determined by 
matching at $r=R$ the form
(\ref{G40a})  and its derivative to the functions (\ref{G40}) and (\ref{G41}), respectively.
This gives two algebraic equations from which $A(\kappa)$ and $B(\kappa)$ are calculated. 
A bound state wave function must be square integrable which implies that
the exploding term proportional to $i_\ell$ 
in (\ref{G40a})  is inadmissible and therefore
the bound  state condition ensuring exponential fall-off,  is $B(\kappa)=0$. Expressing
$B(\kappa)$ in terms of the wave function and its derivative, we obtain
 \begin{equation}
 \phi(\kappa,R)\,k_\ell^\prime(\kappa R) - 
 k_\ell(\kappa R)\,[\chi(\kappa,R)+\phi(\kappa,R)\,\ell/R]=0
 \label{G40b}
 \end{equation}
 where prime denotes derivative with respect to $R$. 
 Solving eq. (\ref{G40b}) for $\kappa$, we are in the position
 to locate the bound states. 
 \par
 When the Coulomb interaction is present the asymptotic form of the wave
 function (\ref{G40a}) needs to be modified, and the appropriate formula reads
 \begin{equation}
 \phi(\kappa,r)\approx (\kappa r)^{-\ell-1}
 \,[A(\kappa)\,\tilde{h}(\kappa r) + B(\kappa)\,\tilde{f}(\kappa r)].
         \label{G40c}
 \end{equation}
 In the case of a bound state we seek the exponentially
 decaying solution  which implies that the  
 term proportional to $\tilde{f}$ must be absent and this provides the 
 condition for the existence of a bound state
 \begin{equation}
         \phi(\kappa,R)\left[ \ell+1 - 
   \kappa R\;\tilde{h}'(\kappa R)/
   \tilde{h}(\kappa R)
  \right ] + R\,\chi(\kappa,R)=0.
         \label{G40d}
 \end{equation} 
 The presence of negative energy Coulomb wave functions
 in (\ref{G40d}) is only a minor complication owing to the fact 
 that all that is needed is the logarithmic derivative of
 the function $\tilde{h}$ given by (\ref{B5f}).  
 The latter combination can be   
 very efficiently computed using a continued fraction representation. 
 \section{Solution of the Lippmann-Schwinger equation}
 The L-S equation is an attractive alternative to the Schr\"{o}dinger equation.
 The method of solution of the L-S equation in the configuration space
 presented in this section will be applied exclusively to
 the Volterra integral equation of the second kind. This is true
 for the continuous and discrete spectrum alike.
 \subsection{Continuous spectrum}
 To calculate the phase shift from (\ref{QM13}), all that is needed is the wave function
 $u(p,r)$ on the Chebyshev mesh in the interval $[0,R]$. The Fourier-Chebyshev interpolation
 is never needed because the integrals entering
 (\ref{QM13}) are computed by applying the Chebyshev quadrature  (\ref{G21}) which
 requires the wave function only at the collocation points.
 The wave function is obtained by solving the Volterra equation (\ref{QM11}) following
 closely the algorithm specified in (\ref{G24}). Introducing the vector $[u]$ containing
 the values of $u(p,x_i)$, we are led to a system of algebraic equations
\begin{equation*}
        \left\{\bm{1}-\Half\,R\,
        \pmb{\mathcal{K}} \circ \bm{W}^- \right\}\cdot [u] = [f],   
\end{equation*}
where the kernel matrix $\pmb{\mathcal{K}}$,
is ${\cal K}_{ij}=(2\mu/p)\,(f_i\,g_j-g_i\,f_j)\,V(x_j)$ and the vectors
$[f]$ and $[g]$ represent the free solutions $f(px_i)$ and $g(px_i)$, respectively.
 \par
 In some applications, to reduce the storage size, it might be necessary to introduce
 a composite integration by chopping
 the integration interval $[0,R]$ into a number of smaller partitions.  
 In order to adapt the above scheme to such a situation we shall
 need a second linearly independent solution of the wave equation 
 that necessarily would be 
 irregular at the origin.  This solution, hereafter denoted as $w(p,r)$, 
 is defined by the  appropriate boundary condition at infinity.
 As a convenient choice we consider the solution of the equation
 \begin{equation}
        w(p,r)=g(pr) 
        +\frac{2\mu}{p} \int_r^\infty \left[ f(pr')\, g(pr) - g(pr')\, f(pr)\right]
        \,V(r')\,w(p,r')\,\D r'. 
        \label{I1}
 \end{equation}   
 The Wronskian $W[w,u]$ does not vanish and, similarly to the free propagation case,
 is equal to $p$. Therefore,  any solution of the wave equation can be written as a
 superposition of the two linearly independent solutions: $u$ and $w$. 
 We wish now to  
 break the interval $[0,R]$ into $M$ partitions $[R_{\lambda-1},R_\lambda]$ with
 $\lambda=1,2,\cdots M$ where $R_0=0$ and $R_M=R$. For greater clarity,  
 we have reserved hereafter a Greek index $\lambda$ to number the partitions
 and letting Roman indices to label the Chebyshev mesh points.
 It is important to distinguish between these two kinds of labels, one
 enumerating the sub-mesh points (dimension $N$) within a given partition,
 and the other enumerating different partitions (dimension $M$). 
 Obviously, the size of the whole mesh is $M\times N$ and, accordingly, the 
 collocation points ought to carry  labels of both types just mentioned.
 Since the Greek index $\lambda$  numbers the partitions
 and Roman indices label the Chebyshev mesh points,
 the collocation
 points are denoted as $x_i^\lambda$ with $\lambda=1,2,\dots,M-1$
 and $i=1,2,\dots,N$, and we have 
 \begin{equation}
         x_i^\lambda=\Half(R_\lambda+R_{\lambda-1})
                    +\Half(R_\lambda-R_{\lambda-1})\,
 \cos[\pi(i-\Half)/N].
         \label{IX6}
 \end{equation}
 In every partition $\lambda=1,2,\dots,M$ we have two linearly independent
 local solutions $u_\lambda$ and $w_\lambda$  satisfying, respectively, the 
 appropriate Volterra equations
 \begin{equation}
        u_\lambda(p,r)=f(pr) 
        -\frac{2\mu}{p} \int_{R_{\lambda-1}}^r 
                \left[ f(pr')\, g(pr) - g(pr') \, f(pr)\right]
        \,V(r')\,u_\lambda(p,r')\,\D r'; 
        \label{I2}
 \end{equation}
 and
  \begin{equation}
        w_\lambda(p,r)=g(pr) 
        +\frac{2\mu}{p} \int_r^{R_\lambda}
                \left[ f(pr') \, g(pr) - g(pr') \, f(pr)\right]
        \,V(r')\,w_\lambda(p,r')\,\D r'; 
        \label{I3}
 \end{equation}     
where it is understood that $r$ belongs to the interval $[R_{\lambda-1},R_\lambda]$.
The general solution in the latter interval 
may be written as $A_\lambda \,u_\lambda (p,r)+B_\lambda \,w_\lambda (p,r)$ 
where $A_\lambda $ and $B_\lambda $ are
two constants to be determined.  Obviously, the global solution constructed from
the local solutions is expected to  be a continuous and smooth function which implies
 that the different 
 local solutions, together with their derivatives,  should be matched at the
partition boundaries $r=R_\lambda$. This gives $2(M-1)$ equations for $2M$  coefficients
$A_\lambda$ and $B_\lambda$  
 \begin{eqnarray*}
 && A_\lambda \,u_\lambda (p,R_\lambda )+B_\lambda \,w_\lambda (p,R_\lambda )=
 A_{\lambda +1}\,u_{\lambda +1}(p,R_\lambda )+B_{\lambda +1}\,w_{\lambda +1}(p,R_\lambda )\\ 
 && A_\lambda \,u_\lambda ^\prime(p,R_\lambda )+B_\lambda \,w_\lambda ^\prime(p,R_\lambda )=
 A_{\lambda +1}\,u_{\lambda +1}^\prime(p,R_\lambda )+
 B_{\lambda +1}\,w_{\lambda +1}^\prime(p,R_\lambda ),
 \end{eqnarray*} 
with $\lambda=1,2,\dots,M$, resulting in two recurrences
\begin{eqnarray}
&&A_{\lambda+1}=  a_\lambda\;A_\lambda + b_\lambda\;B_\lambda;
\label{I4}\\
&&B_{\lambda+1}=\alpha_\lambda\;A_\lambda + \beta_\lambda\;B_\lambda
\label{I5}
\end{eqnarray} 
with
\begin{eqnarray*}
    a_\lambda   &=&[u_\lambda (p,R_\lambda )  \, w^\prime_{\lambda +1}(p,R_\lambda ) - u^\prime_\lambda (p,R_\lambda )\,       w_{\lambda +1}(p,R_\lambda )]/d;\\
    b_\lambda   &=&[w_\lambda (p,R_\lambda )  \, w^\prime_{\lambda +1}(p,R_\lambda ) - w^\prime_\lambda (p,R_\lambda )\,       w_{\lambda +1}(p,R_\lambda )]/d;\\
\alpha_\lambda  &=&[u_\lambda ^\prime(p,R_\lambda ) \,  u_{\lambda +1}(p,R_\lambda ) -        u_\lambda (p,R_\lambda )\,u^\prime_{\lambda +1}(p,R_\lambda )]/d;\\
\beta_\lambda   &=&[w_\lambda ^\prime(p,R_\lambda ) \,  u_{\lambda +1}(p,R_\lambda ) -
w_\lambda (p,R_\lambda )\,u^\prime_{\lambda +1}(p,R_\lambda )]/d;\\ 
      d  &=& u_{\lambda +1}(p,R_\lambda )\,w^\prime_{\lambda +1}(p,R_\lambda ) -
          u^\prime_{\lambda +1}(p,R_\lambda )\,w_{\lambda +1}(p,R_\lambda ).
\end{eqnarray*}
 The above expression involve the derivatives of the two local solutions
 $u_\lambda$ and $w_\lambda$ which are obtained
 by differentiating  (\ref{I2}) and (\ref{I3}). Although in
 both cases the integrand vanishes for $r^\prime=r$, but there would be a contribution from
 the derivative of the kernel. One has
   \begin{equation*}
        u_\lambda ^\prime(p,r)=f'(pr) 
                -\dfrac{2\mu}{p} \int_{R_{\lambda -1}}^r 
                \left[ f(pr') \, g'(pr) - g(pr') \, f'(pr)\right]
        \,V(r')\,u_\lambda (p,r')\,\D r'; 
   \end{equation*} 
and
  \begin{equation*}
        w_\lambda ^\prime(p,r)= g'(pr) 
                +\dfrac{2\mu}{p} \int_r^{R_\lambda } 
                \left[ f(pr') \, g'(pr) - g(pr') \, f'(pr)\right]
        \,V(r')\, w_\lambda (p,r')\,\D r'; 
\end{equation*} 
 where prime denotes derivative with respect to $r$.
 \par
 Since the global solution must be regular,  the starting values are
 $A_1=1$ and $B_1=0$ and this allows to solve the recurrence for
all the remaining coefficients.  The  phase shift is obtained by matching the 
global solution and its derivative at $r=R$ to the asymptotic form
 $ A[f(pr) + \tan\delta\,g(pr)]$. As a result,
 the phase shift would be obtained from (\ref{G36e}) putting
 $\psi(p,R)=A_M\,u_M(p,R)+B_M\,w_M(p,R)$.
\subsection{Bound states}
To locate the bound states we need to calculate the Fredholm determinant 
$\Delta(\kappa)$ from (\ref{B6}). This, in turn, requires the solution of 
of the Volterra equation of the second kind (\ref{B7}). 
 This equation is solved for
 the wave function  $\tilde{u}(\kappa,r)$ on the Chebyshev mesh using
 the algorithm  (\ref{G24}). Introducing the vector $[\tilde{u}]$ containing
 the values of $\tilde{u}(\kappa,x_i)$, we are led to a system of algebraic 
 equations
\begin{equation*}
        \left\{\bm{1}- \Half R\,
        \Calbf{K} \circ \bm{W}^- \right\}\cdot[\tilde{u}]=[\tilde{f}],   
\end{equation*}
where the kernel $\Calbf{K}$, is 
${\cal K}_{ij}= (2\mu/\kappa)\, (\tilde{f}_i\,\tilde{h}_j-\tilde{h}_i\,\tilde{f}_j)\,V(x_j)$ 
and the arrays
$[\tilde{f}]$ and $[\tilde{h}]$ represent, respectively, the grid values of the free  
wave functions $\tilde{f}(\kappa x_i)$ and $\tilde{h}(\kappa x_i)$.
With the solution $[\tilde{u}]$ in hand, the Fredholm determinant, is
given as
\begin{displaymath}
        \Delta(\kappa)=1 + \dfrac{R\mu}{\kappa}\,\displaystyle\sum_{i=1}^N\,\tilde{h}_i\,
        V(x_i)\,\tilde{u}_i\,\tilde{w}_i.
\end{displaymath}
Finally,
the bound states would be determined by finding the zeros of  $\Delta(\kappa)$.
\par
If need arises, composite integration algorithm
may also be applied, and the procedure is very  
similar to that used in the continuous spectrum case. Thus,
the $[0,R]$ range is divided into $M$ partitions $R_1,\,R_2,\dots,\,R_M$
and in each partition $\lambda$ we have two linearly independent wave functions 
$\tilde{u}_\lambda$ and $\tilde{w}_\lambda$. These wave functions are obtained by solving
the appropriate Volterra equations 
 \begin{equation}
         \tilde{u}_\lambda (\kappa,r)=\tilde{f}(\kappa r) 
         -\frac{2\mu}{\kappa} \int_{R_{\lambda -1}}^r 
                \left[ \tilde{f}(\kappa r')\, \tilde{h}(\kappa r) - \tilde{h}(\kappa r') \,
                \tilde{f}(\kappa r)\right]
                \,V(r')\,\tilde{u}_\lambda (\kappa,r')\,\D r'; 
        \label{I2a}
\end{equation}
and
  \begin{equation}
          \tilde{w}_\lambda (\kappa,r)=\tilde{h}(\kappa r) 
        +\frac{2\mu}{\kappa} \int_r^{R_\lambda }
                \left[ \tilde{f}(\kappa r') \, 
                \tilde{h}(\kappa r) - \tilde{h}(\kappa r') \, \tilde{f}(\kappa r)\right]
                \,V(r')\,\tilde{w}_\lambda (\kappa,r')\,\D r'; 
        \label{I3a}
\end{equation}     
where $r$ is in the interval $[R_{\lambda-1},R_\lambda]$ with $R_0=0$ and $R_M=R$. The wave
function would be a superposition of the two linearly independent local solutions  
\begin{displaymath}
        \tilde{u}(\kappa,r)=\tilde{A}_\lambda\,\tilde{u}_\lambda(\kappa,r)+
        \tilde{B}_\lambda\,\tilde{w}_\lambda(\kappa,r).
\end{displaymath}
The coefficients $\tilde{A}_\lambda$ and $\tilde{B}_\lambda$ can be determined by 
smoothly matching
 the different local pieces of $\tilde{u}$  and its derivative with respect to $r$
 at all partition boundaries. The matching conditions, are 
 \begin{eqnarray*}
        && \tilde{A}_\lambda \,\tilde{u}_\lambda (\kappa,R_\lambda )+\tilde{B}_\lambda \,\tilde{w}_\lambda (\kappa,R_\lambda )=
  \tilde{A}_{\lambda +1}\,\tilde{u}_{\lambda +1}(\kappa,R_\lambda )+
  \tilde{B}_{\lambda +1}\,\tilde{w}_{\lambda +1}(\kappa,R_\lambda )\\ 
        && \tilde{A}_\lambda \,\tilde{u}_\lambda ^\prime(\kappa,R_\lambda )+\tilde{B}_\lambda \,\tilde{w}_\lambda ^\prime(\kappa,R_\lambda )= 
        \tilde{A}_{\lambda +1}\,\tilde{u}_{\lambda +1}^\prime(\kappa,R_\lambda )+
        \tilde{B}_{\lambda +1}\,\tilde{w}_{\lambda +1}^\prime(\kappa,R_\lambda ),
 \end{eqnarray*} 
 and yield the following recurrence relations for the coefficients
\begin{eqnarray}
        &&\tilde{A}_{\lambda +1}=  a_\lambda \;\tilde{A}_\lambda  + b_\lambda \;\tilde{B}_\lambda ;
\label{I4a}\\
&&\tilde{B}_{\lambda +1}=\alpha_\lambda \;\tilde{A}_\lambda  + \beta_\lambda \;\tilde{B}_\lambda 
\label{I5a}
\end{eqnarray} 
where
\begin{eqnarray*}
   && a_\lambda   =[\tilde{u}_\lambda (\kappa,R_\lambda )  \, 
   \tilde{w}^\prime_{\lambda +1}(\kappa,R_\lambda ) - 
        \tilde{u}^\prime_\lambda (\kappa,R_\lambda )\,     
        \tilde{w}_{\lambda +1}(\kappa,R_\lambda )]/d;\\
   && b_\lambda   =[\tilde{w}_\lambda (\kappa,R_\lambda ) 
   \, \tilde{w}^\prime_{\lambda +1}(\kappa,R_\lambda ) -
        \tilde{w}^\prime_\lambda (\kappa,R_\lambda )\,   
        \tilde{w}_{\lambda +1}(\kappa,R_\lambda )]/d;\\
   &&\alpha_\lambda  =[\tilde{u}_\lambda ^\prime(\kappa,R_\lambda ) \, 
   \tilde{u}_{\lambda +1}(\kappa,R_\lambda ) -
   \tilde{u}_\lambda (\kappa,R_\lambda )\,\tilde{u}^\prime_{\lambda +1}(\kappa,R_\lambda )]/d;\\
   &&\beta_\lambda   =[\tilde{w}_\lambda ^\prime(\kappa,R_\lambda ) \,
   \tilde{u}_{\lambda +1}(\kappa,R_\lambda ) -
   \tilde{w}_\lambda (\kappa,R_\lambda )\,\tilde{u}^\prime_{\lambda +1}(\kappa,R_\lambda )]/d;\\ 
   &&   d  = \tilde{u}_{\lambda +1}(\kappa,R_\lambda )\,
   \tilde{w}^\prime_{\lambda +1}(\kappa,R_\lambda ) -
   \tilde{u}^\prime_{\lambda +1}(\kappa,R_\lambda )\,\tilde{w}_{\lambda +1}(\kappa,R_\lambda ).
\end{eqnarray*}
The derivatives occurring in the above expressions are obtained by differentiation
of the Volterra equations (\ref{I2a}) and (\ref{I3a}), and we get
   \begin{eqnarray}
         &&  \tilde{u}_\lambda ^\prime(\kappa ,r)=\tilde{f}'(\kappa r) 
                -\dfrac{2\mu}{\kappa } \int_{R_{\lambda -1}}^r 
                \left[ \tilde{f}(\kappa r') \, \tilde{h}'(\kappa r) - 
                \tilde{h}(\kappa r') \, \tilde{f}'(\kappa r)\right]
        \,V(r')\,\tilde{u}_\lambda (\kappa ,r')\,\D r'; 
                \label{IX3} \\
         && \tilde{w}_\lambda ^\prime(\kappa ,r)= \tilde{h}'(\kappa r) 
                +\dfrac{2\mu}{\kappa } \int_r^{R_\lambda } 
                \left[ \tilde{f}(\kappa r') \, \tilde{h}'(\kappa r) -
                \tilde{h}(\kappa r') \, \tilde{f}'(\kappa r)\right]
        \,V(r')\,\tilde{w}_\lambda (\kappa ,r')\,\D r'. 
                \label{IX4} 
\end{eqnarray} 
The bound state function must be regular which implies that $A_1=1$ and $B_1=0$
are the proper starting values in   (\ref{I4a}) and (\ref{I5a}). 
The wave function in the asymptotic region $r\geq R$, has the form
\begin{equation}
        \tilde{u}(\kappa,r)=\tilde{A}(\kappa)\,\tilde{f}(\kappa r)+
        \tilde{B}(\kappa)\,\tilde{h}(\kappa r)
        \label{IX1}
\end{equation}
where $\tilde{A}(\kappa)$ and  $\tilde{B}(\kappa)$ are obtained by matching 
at $r=R$ the asymptotic solution (\ref{IX1}) to the wave function from the last partition
and subsequently repeating this procedure for the derivatives. 
For arbitrary value of $\kappa$ the wave function (\ref{IX1}) would not be
square integrable owing to the presence of exponentially increasing function
$\tilde{f}(\kappa,r)$. However, for some particular $\kappa$ values for which
$\tilde{A}(\kappa)=0$ the exploding term is absent and the wave function shows exponential
fall-off for large $r$. For such $\kappa$ we have a bound state. The ultimate
bound state condition is obtained by calculating $\tilde{A}(\kappa)$ from
the matching condition and equating it to zero. This gives
\begin{equation}
        [\tilde{A}_M\,\tilde{u}_M(\kappa,R)+\tilde{B}_M\,\tilde{w}_M(\kappa,R)]
        \,\tilde{h}^\prime(\kappa R) -
        [\tilde{A}_M\,\tilde{u}_M^\prime(\kappa,R)+ \tilde{B}_M\,\tilde{w}_M^\prime(\kappa,R)]
        \,\tilde{h}(\kappa R)=0.
        \label{IX2}
\end{equation}
 The bound state algorithm is implemented as follows.  
 Firstly, for each of the $M$
 partitions the two sets of $N$ algebraic equations must be solved
 \begin{eqnarray*}
        && \left\{\bm{1}-\Half(R_\lambda-R_{\lambda-1})\;
  \Calbf{K}_\lambda \circ \bm{W}^- \right\}\cdot[\tilde{u_\lambda}]=[\tilde{f_\lambda}], \\
        && \left\{\bm{1}+\Half(R_\lambda-R_{\lambda-1})\;
  \Calbf{K}_\lambda \circ \bm{W}^+ \right\}\cdot[\tilde{w_\lambda}]=[\tilde{h_\lambda}].   
 \end{eqnarray*}
 where for fixed $\lambda$ the kernel $\Calbf{K}_\lambda$ is a $N\times N$ matrix  
 \begin{displaymath}
         [\Calbf{K}_\lambda]_{ij}=
 (2\mu/\kappa)\;[\tilde{f}(\kappa\,x_i^\lambda)\,\tilde{h}(\kappa\,x_j^\lambda)-
  \tilde{h}(\kappa\,x_i^\lambda)\,\tilde{f}(\kappa\,x_j^\lambda)]\;V(x_j^\lambda) 
 \end{displaymath}
 and the arrays
$[\tilde{f}_\lambda]$ and $[\tilde{h}_\lambda]$ represent, 
respectively, the grid values of the free  
wave functions $\tilde{f}(\kappa\,x_i^\lambda)$ and $\tilde{h}(\kappa\,x_i^\lambda)$
with $i,j=1,2,\dots,N$.  The solutions at the collocation points supply enough 
information to reconstruct the wave functions locally by Chebyshev interpolation,
and in particular at the matching points $R_{\lambda-1}$ and $R_\lambda$. 
\par
Knowing the local functions $\tilde{u}_\lambda(\kappa, r)$ and 
$\tilde{w}_\lambda(\kappa, r)$ at the collocation points that belong to the partition $\lambda$,
we are also in the position to compute the $r$-derivatives
 of these functions with the aid of (\ref{IX3}) and (\ref{IX4}) by performing
the Chebyshev integrations. The derivative arrays, are  
\begin{eqnarray*}
 &&[\tilde{u}_\lambda^\prime]=[\tilde{f}_\lambda]-\Calbf{K}^\prime_\lambda\circ\bm{W}^+
   \cdot[\tilde{u}_\lambda]\\
 &&[\tilde{w}_\lambda^\prime]=[\tilde{h}_\lambda]+\Calbf{K}^\prime_\lambda\circ\bm{W}^+
  \cdot [\tilde{w}_\lambda]
 \end{eqnarray*}
 with the derivative kernel given as
 \begin{displaymath}
         [\Calbf{K}^\prime_\lambda]_{ij}=
 (2\mu/\kappa)\;[\tilde{f}(\kappa\,x_i^\lambda)\,\tilde{h}^\prime(\kappa\,x_j^\lambda)-
  \tilde{h}(\kappa\,x_i^\lambda)\,\tilde{f}^\prime(\kappa\,x_j^\lambda)]\;V(x_j^\lambda) 
 \end{displaymath}
 where prime denotes the derivative with respect to $r$. 
 \section{Momentum space solutions}
 \subsection{Continuous spectrum}
 In order to solve (\ref{sing9}), we take a simple rational transformation
 \begin{equation}
 q=p\sigma(1+t)/(1-t)
         \label{sing17}
 \end{equation}
  mapping the infinite integration 
  range onto the $[-1,1]$ interval. 
 The presence of
 an adjustable slope parameter $\sigma$   
 in (\ref{sing17}) improves the flexibility of this transformation. 
 Now, invoking (\ref{G2B}), 
 both, the potential and the off-shell K-matrix, will be expressed
 in terms of the Chebyshev cardinal functions. For the
 potential we just use the interpolating formula
 \begin{equation}
         U_\ell(k',k'')=\sum_{i,j=1}^N G_i(t')\,U_{ij}\,G_j(t'')                 
         \label{sing18}
 \end{equation}
 with $U_{ij}=U_\ell(k_i,k_j)$ and
 $k_j= p\sigma(1+t_j)/(1-t_j)$.
 Similar expansion is used for the off-shell K-matrix 
 \begin{equation}
  \bra{k'}K_\ell(p^2)\ket{k''} =          
         \sum_{i,j=1}^N G_i(t')\,K_{ij}(p^2)\,G_j(t'')           
         \label{sing19}
 \end{equation}
 where in both cases above the rational mapping (\ref{sing17})
 provides the relation connecting $k'$ with $t'$ and $k''$ with $t''$, respectively.
 The hitherto unknown matrix  $K_{ij}(p^2)$ 
 will be determined by solving (\ref{sing9}). Using the new
 variables  and the expansions (\ref{sing18}) and (\ref{sing19}),
 the L-S equation takes the matrix form
 \begin{equation}
 \left(\bm{1}+\bm{U}\cdot\bm{\Gamma}\right)\cdot\bm{K}(p^2)=\bm{U}       
         \label{sing20}
 \end{equation}
 where the matrix $\mathbf{\Gamma}$ is given as the integral
 \begin{equation}
         \Gamma_{ij}(p)=     
         \dfrac{4p\sigma^3}{\pi(\sigma+1)^2}
         \int_{-1}^1 
 \dfrac{G_i(t)\,G_j(t)}{(t-\tau)(1-t\tau)} \left(\dfrac{1+t}{1-t}\right)^2
 \; \D t
         \label{sing21}
 \end{equation}
 with $\tau = (1-\sigma)/(1+\sigma)$.
 The integral occurring in (\ref{sing21}) will be evaluated
 using the automatic quadrature (\ref{sing1}). 
 The presence of the cardinal functions in the integrand
 brings major simplifications because in the summation
 over the Gauss-Chebyshev abscissas only a single term survives 
 and the resulting matrix $\mathbf{\Gamma}$ is diagonal
 \begin{displaymath}
 \bm{\Gamma}_{ij}= 
 \delta_{ij}
 \; 4p\sigma^3/[\pi(1+\sigma)^2]
 \; \omega_i(\tau)/[1-t_i\tau]\;
 \left[(1+t_i)/(1-t_i)\right]^2.
 \end{displaymath}
 The explicit solution of (\ref{sing20}) 
 \begin{equation}
 \bm{K}(p^2)=    
         \left(\bm{1}+\bm{U}\cdot\bm{\Gamma}\right)^{-1}\cdot\bm{U}      
         \label{sing23}
 \end{equation}
 inserted in (\ref{sing19}), yields the desired off-shell K-matrix. 
 \par
 It is evident from (\ref{sing17}) that the 
  mapping introduces a singularity at $t=1$ 
  reflecting the fact that the original integral was in the
 infinite range. The singular point $t=1$ corresponds to $q$ going
 to infinity in (\ref{sing9}) but, in all cases of physical interest,
 the potential goes to zero  much faster than quadratically
 when this limit is taken, making the
 $t=1$ singularity quite harmless.  
 \par
 When only the phase shift is needed it is sufficient to calculate
 the half-off-shell K-matrix $ \bra{k_j}K_\ell(p^2)\ket{p}$ on the
 Chebyshev mesh  $k_j, \;j=1,2,\dots N$. The on-shell K-matrix
 is subsequently obtained, from the expansion
 \begin{displaymath}
  \bra{p}K_\ell(p^2)\ket{p} =
    \sum_{j=1}^N G_j(\tau)\; \bra{k_j}K_\ell(p^2)\ket{p},
 \end{displaymath}
 where 
 \begin{displaymath}
         \bra{k_j}K_\ell(p^2)\ket{p}=\sum_{i=1}^N
         \left(\bm{1}+\bm{U}\cdot\bm{\Gamma}\right)^{-1}_{ji}\;
         U(k_i,p).       
 \end{displaymath}
 The linear system (\ref{sing20}) needs to be solved only once
 to get the on-shell K-matrix whereas to determine the fully off-shell
 K-matrix this operation must be repeated $N$ times.
 \subsection{Bound states}
 In momentum space the Schr\"odringer and the L-S equation both result in
 the same scheme for locating the bound states. We begin with the
 Schr\"odinger equation and we wish to apply the 
 Gauss-Chebyshev quadrature which requires that
 the $[0,\infty]$ integration range to be mapped onto $[-1,1]$ interval.   
 This may be accomplished by a change of the integration variable
 \begin{equation}
         k'=(\sigma/a)\,(1+t)/(1-t) 
         \label{sing25}
 \end{equation}
 where $a$ is the range of the potential and $\sigma$ is an adjustable slope parameter.
 The integration over $t$ is then carried out by the automatic quadrature (\ref{G12})
 and we end up with a real and symmetric eigenvalue problem
 \begin{equation}
         [\bm{1} - \lambda(\kappa)\;\pmb{\mathcal{M}}(\kappa^2)] \cdot \tilde{u} = 0  
 \label{sing25a}
 \end{equation}
 which can be solved by standard methods.  
 The explicit form of the matrix $\pmb{\mathcal{M}}$, is
 \begin{equation*}
         \mathcal{M}_{ij}(\kappa^2)=-(4\sigma/a\pi)\;
  [\sqrt{w_i}/(1-t_i)]\;
  (k_i/\sqrt{\kappa^2+ k_i^2}) 
 \; U_\ell(k_i, k_j)\;
  (k_j/\sqrt{\kappa^2+ k_j^2}) 
  \; [ \sqrt{w_j} /(1-t_j)]  
 \end{equation*}
 where the new wave function is $\tilde{u}_j= [\sqrt{w_j}/(1-t_j)]\,u(k_j)$.
 Solving the algebraic eigenvalue problem repeatedly
 for a number of $\kappa$ values, we
 get in each case the value of $\lambda(\kappa)$.  When the latter function
 has been tabulated, it may  be locally approximated by polynomials and
 it remains to apply inverse interpolation to solve
 the equation $\lambda(\kappa_0)=1$.
 \par
 The above procedure applies also for Coulomb potential but this
 case requires some care.
 The point charge Coulomb potential in momentum space $U_\ell(k, k')$
 exhibits in all partial waves a logarithmic singularity
 at $k'=k$ (cf. Appendix).
 As a result, the diagonal elements of the $\pmb{\mathcal{M}}$ matrix
 become infinite and the momentum space framework is in trouble.
 In the literature, this difficulty is alleviated by a sophisticated 
 subtraction scheme serving to eliminate the singularity \cite{tabakin}\cite{heddle}. 
 Within the Chebyshev approach, however, the Coulomb problem 
 in momentum space can be solved directly without subtractions.
 In order to achieve this goal 
 the integrals involving the logarithmic singularity 
 need to be approximated by the dedicated quadrature (\ref{sing13}). 
 \par
 It might be instructive to dwell on the
 momentum space hydrogen-like problem.
 In this case, we are presented with the integral equation
 (for the notation cf. Appendix)
 \begin{equation}
   u(\xi) = \dfrac{2}{\pi} \int_0^\infty 
   \left\{P_\ell(z)\,\log(\xi'+\xi)-W_{\ell-1}(z)
   -P_\ell(z)\,\log|\xi'-\xi|\right\} \;
   \dfrac{u(\xi')\,\D \xi'}{ \sqrt{ (x^2+\xi^2) (x^2+\xi^{'2}) } }.      
         \label{Coul1}
 \end{equation}
 In (\ref{Coul1}) we switched to  dimensionless variables, and
 $\xi, \xi'$ and $x$ are defined, respectively, as $ka, k'a$ and $\kappa a$ where $a$ is the 
 appropriate Bohr radius.  The integrals in (\ref{Coul1})  containing the non-singular terms
 can be approximated by the Gauss-Chebyshev quadrature, whereas the last integral
 involving the  singular $\log|\xi'-\xi|$ term requires the dedicated quadrature (\ref{sing13}).
 Changing the integration variable
 \begin{equation}
          \xi=\sigma \, (1+t)/(1-t)
         \label{Coul0}
 \end{equation}
 the resulting matrix $\pmb{\mathcal{M}}$, takes the form
 \begin{equation}
         \mathcal{M}_{ij}=\dfrac{4\sigma}{\pi}\;
         \dfrac{w_j\,\left\{P_\ell(z_{ij})\,\log|1-t_i\,t_j|-W_{\ell-1}(z_{ij})\right\}
  -P_\ell(z_{ij})\,\Omega_j(t_i)}
  {(1-t_j)^2\,\sqrt{(\xi^2_i+x^2)(\xi^2_j+x^2)} },
         \label{Coul2}
 \end{equation}
 where $z_{ij}=(\xi_i^2+\xi_j^2)/(2\xi_i\xi_j)$
 and it is apparent that the diagonal terms are all finite and well defined.
 Since the matrix $\pmb{\mathcal{M}}$ in (\ref{Coul2}) is not symmetric anymore 
 the eigenvalues have to be determined by solving the secular equation 
 $\det|\bm{1}-\pmb{\mathcal{M}}(x^2)|=0$.
 The same procedure as above applies also to other problems in which the point charge
 Coulomb potential is supplemented by an additional short range potential. 
 \par
 We turn now to the approach based on the L-S equation.
 When the momentum $p$ is on the imaginary axis, we
 set as before $p=\I\,\kappa$ and the L-S equation for the T-matrix, reads 
  \begin{equation}
  \bra{k'}T_\ell(\kappa^2)\ket{k}=U_\ell(k',k)-\dfrac{2}{\pi}\int_0^\infty        
  U_\ell(k',q) \; \dfrac{q^2\,\D q}{q^2+\kappa^2}\; 
  \bra{q}T_\ell(\kappa^2)\ket{k}
          \label{sing24}
  \end{equation}
 and the integral equation is non-singular.   
 We are going to use again the same mapping (\ref{sing25}).
 Expanding the potential
 and the T-matrix in the Chebyshev cardinal functions, we set
 \begin{equation}
  \bra{k'}T_\ell(\kappa^2)\ket{k''} =     
         \sum_{i,j=1}^N G_i(t')\,T_{ij}(\kappa^2)\,G_j(t'')              
         \label{sing26}
 \end{equation}
 and the L-S equation (\ref{sing24}) takes the matrix form
 \begin{equation}
         \mathbf{T}(\kappa^2)=\mathbf{U}-
         \mathbf{U}\cdot\mathbf{\Gamma}(\kappa^2)\cdot\mathbf{T}(\kappa^2)
         \label{sing27}
 \end{equation}
 where the matrix $\bm{\Gamma}(\kappa^2)$, is
 \begin{equation}
  \Gamma_{ij}(\kappa^2) =        
         \dfrac{4\sigma}{\pi a}
         \; \int_{-1}^1 
 \dfrac{ G_i(t)\,G_j(t) } { 1+(\kappa a/\sigma)^2 (1-t)^2/(1+t)^2 } 
  \; \dfrac{\D t}{(1-t)^2}.
         \label{sing28}
 \end{equation}
 Applying  Gauss-Chebyshev quadrature, we obtain
 \begin{equation}
         \Gamma_{ij}(\kappa^2) =         \delta_{ij}\;
         (4\sigma/\pi a)
  \; w_j/(1-t_j)^2
  /[ 1+(\kappa a/\sigma)^2 (1-t_j)^2/(1+t_j)^2 ] 
         \label{sing29}
 \end{equation}
 and it is evident that the matrix $\mathbf{\Gamma}$ is positive definite.
 Therefore, we may dissolve $\mathbf{\Gamma}$ as
 $\mathbf{\Gamma}^{\Half} \cdot \mathbf{\Gamma}^{\Half}$ 
  in (\ref{sing27}) and defining a new t-matrix 
  $\pmb{\mathcal{T}}=\mathbf{\Gamma}^{\Half}\cdot\mathbf{T}\cdot\mathbf{\Gamma}^{\Half}$,
  the L-S equation, becomes
  \begin{equation}
        \left\{\bf{1}-\Calbf{M}(\kappa^2)\right\}\cdot
        \pmb{\mathcal{T}}(\kappa^2)   
        =-\pmb{\mathcal{M}}(\kappa^2),   
          \label{sing29a}
  \end{equation}
 where    $\pmb{\mathcal{M}}=-\mathbf{\Gamma}^{\Half}\cdot\mathbf{U}\cdot\mathbf{\Gamma}^{\Half}$.
 Since the calculation of the t-matrix involves inverting the 
 matrix $\bm{1}-\pmb{\mathcal{M}}(\kappa^2)$, the poles of 
 $T$ will occur at such values of $\kappa$ at which the determinant
 of the latter matrix vanishes. Thus, the ultimate equation
 for $\kappa$, takes the form
 \begin{equation}
 \text{Det}|\bm{1}-\pmb{\mathcal{M}}(\kappa^2)|=0.
         \label{sing30}
 \end{equation}
 Clearly, (\ref{sing30}) is nothing else but the secular equation associated
 with the eigenvalue problem (\ref{sing25a}) 
 which demonstrates the equivalence of L-S and
 Schr\"odinger approach.
 \par
 The above scheme can be also used to calculate the scattering length $A$.
 From (\ref{sing26}), we obtain 
 \begin{equation}
         A =-\bra{0}T_\ell(0)\ket{0} =    
         \sum_{i,j=1}^N G_i(-1)\,T_{ij}(0)\,G_j(-1)              
         \label{sing31}
 \end{equation}
 with
 \begin{equation}
 \bm{T}(0)=
 [\bm{1}+\bm{U}\cdot\bm{\Gamma}(0)]^{-1}\cdot \bm{U}. 
         \label{sing32}
 \end{equation}
 The matrix $\bm{\Gamma}(0)$ is diagonal and takes a simple form
 \begin{equation}
         \Gamma_{ij}(0)=\delta_{ij}\,(4s/\pi)\;[\sigma\, w_j/(1-t_j)^2]. 
         \label{sing33}
 \end{equation}
 Summarizing, 
 the scattering length is given as a scalar product of two vectors
 \begin{equation}
         A=-[G]\cdot [X]
         \label{sing34}
 \end{equation}
 where the cardinal functions in the array $[G]$ are to be taken at $t=-1$ and 
 the vector $[X]$ is obtained as the solution of a linear system of equations
 \begin{equation}
         \{\bm{1}+\bm{U}\cdot\bm{\Gamma}(0)\}\cdot[X]=\bm{U}\cdot[G].
         \label{sing35}
 \end{equation}
\section{Numerical performance test}
 Having set up the theoretical background, we are ready now to investigate
 the performance of the semi-spectral Chebyshev method by solving numerically
 some concrete differential and integral equations 
 encountered in quantum mechanics and subsequently comparing the results with the
 exact solutions. As mentioned before, we confine our attention to time independent
 non-relativistic single-channel two-body problems, 
 disregarding all internal degrees of freedom.    
 We considered three popular potential shapes: (i) exponential, (ii) Hulth\'{e}n
 and (iii) Morse for which exact solutions are available and have been
 collected in the Appendix.  We solved numerically the s-wave scattering problem
 and located the bound states for these potentials by solving three different equations:
 (i) the differential equation (Schr\"{o}dinger), (ii) the Volterra integral
 equation (L-S in configuration space), and (iii) the Fredholm integral equation
 (L-S in momentum space). Finally, we solved numerically the bound state problem
 for a point-charge Coulomb potential, both in the configuration and in 
 the momentum space.
 \par
\begin{figure}
 \centering      
 \begin{minipage}[t]{.4\textwidth}
 \centering
  \includegraphics[scale=0.35]{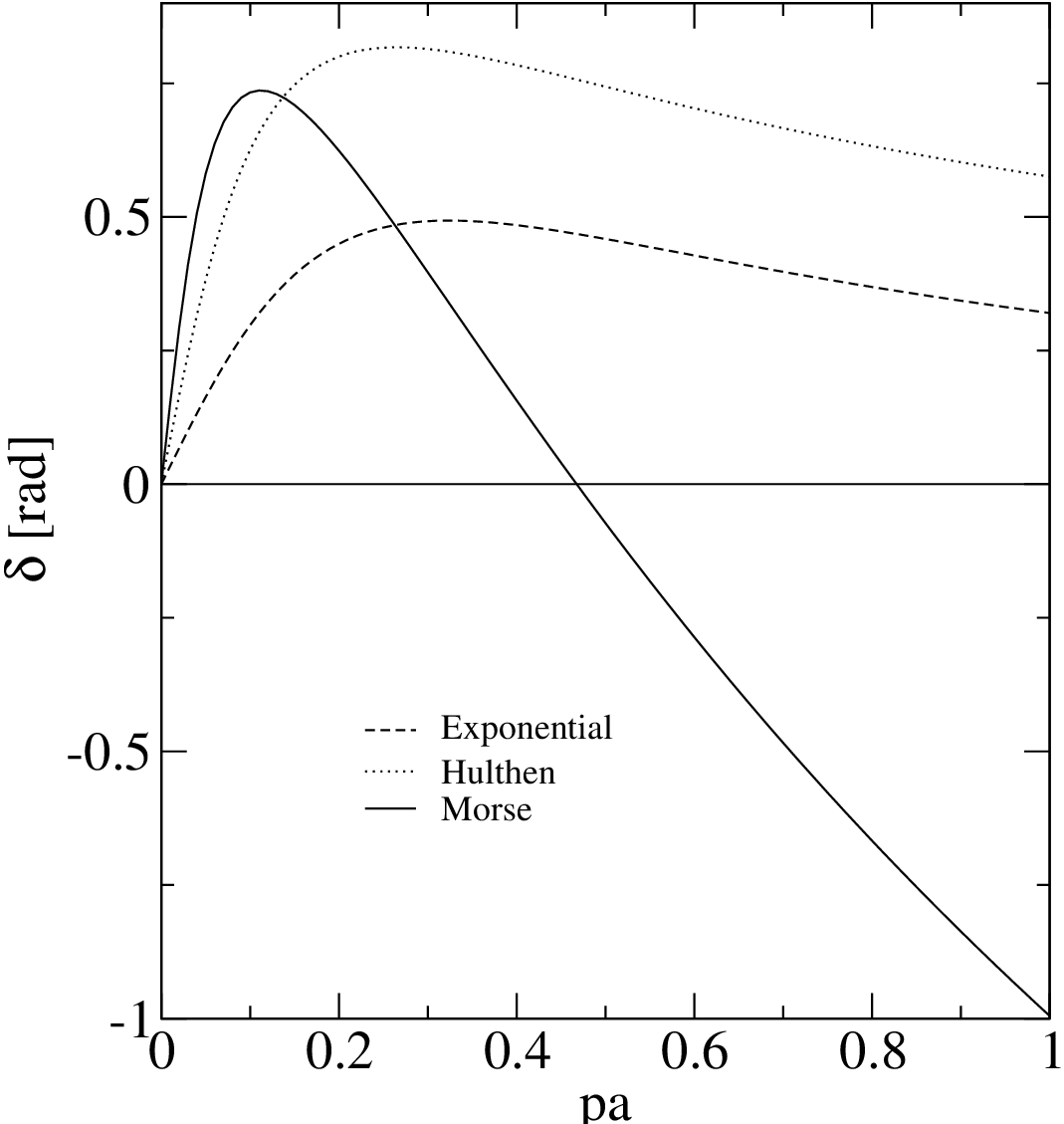}
       \caption{Phase-shift (L=0) for different potentials
       vs.  momentum in inverse range units. }
     \label{fig:one}
   \end{minipage}
   \hspace{1cm}
   \begin{minipage}[t]{.4\textwidth}
   \centering
   \includegraphics[scale=0.35]{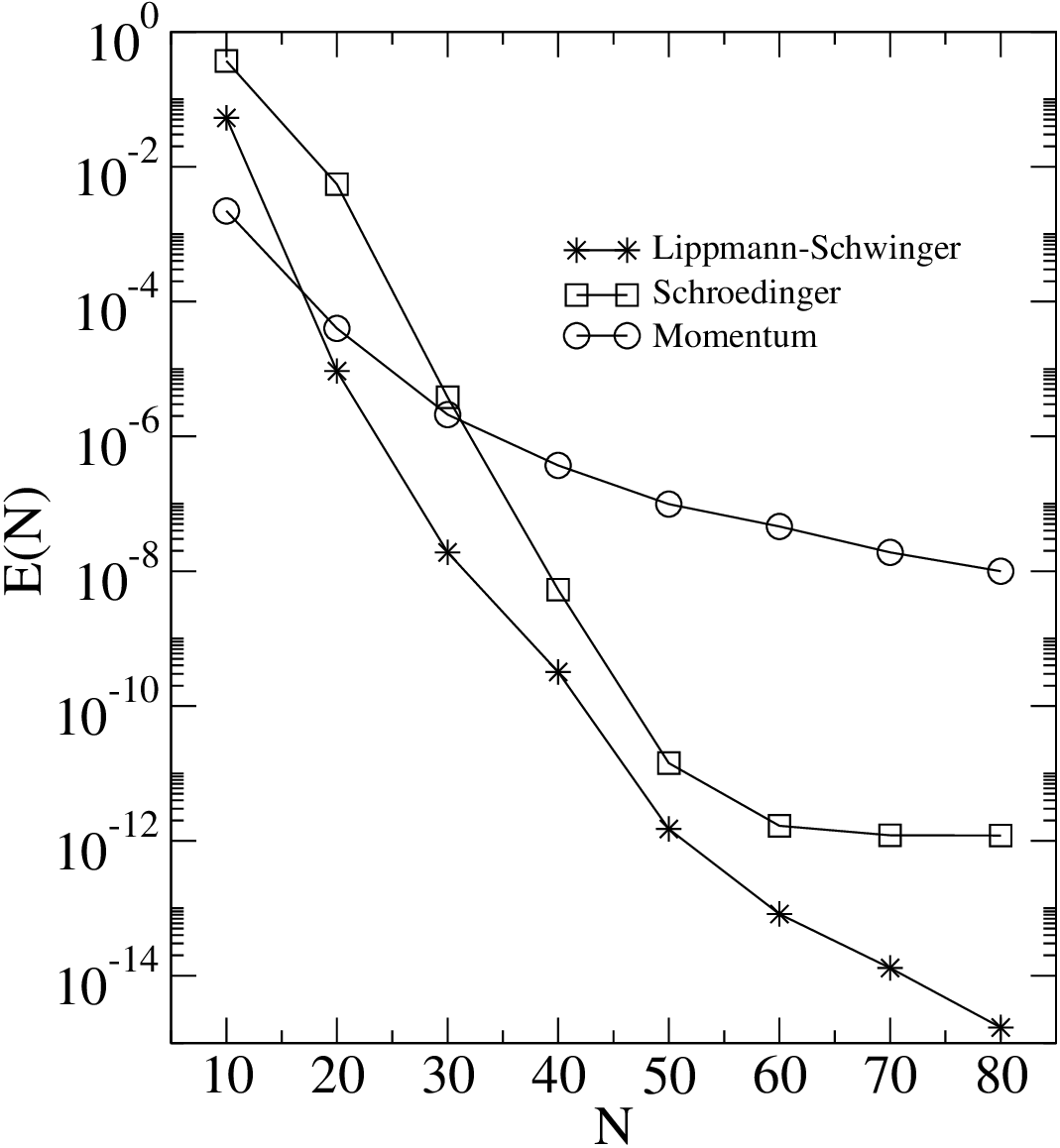}
   \caption{
   Average relative error (\ref{result1})
   for the  phase-shift
   versus $N$  (exponential potential).}
  \label{fig:two}
  \end{minipage}
  \vspace*{1.5cm}
  \end{figure}
\begin{figure}
  \vspace*{1cm}	
 \centering      
 \begin{minipage}[b]{.4\textwidth}
 \centering
  \includegraphics[scale=0.35]{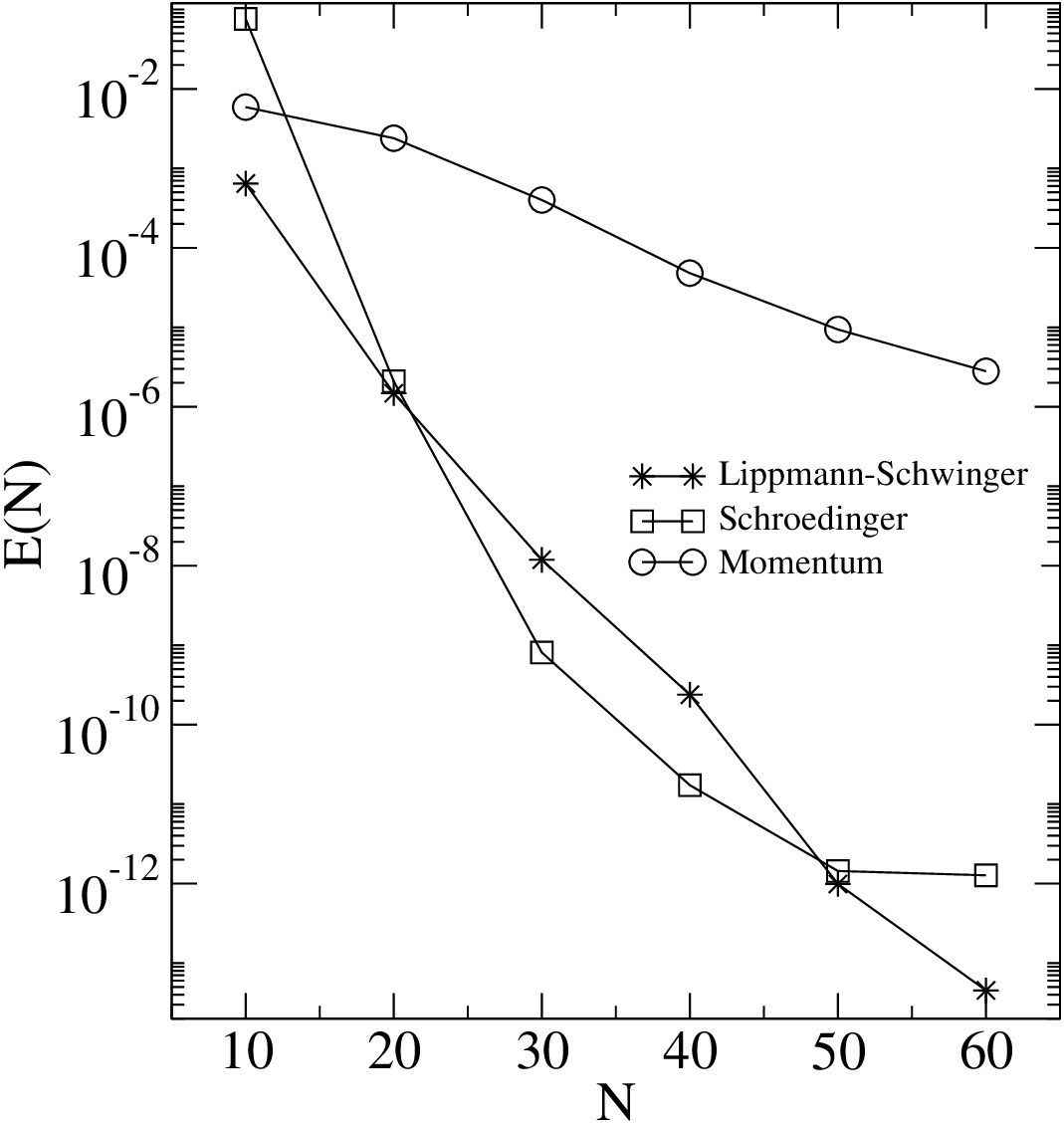}
   \caption{
   Average relative error (\ref{result1})
   for the  phase-shift
   versus $N$  (Hulth\'{e}n potential).}
     \label{fig:three}
   \end{minipage}
    \hspace{1cm}
   \begin{minipage}[b]{.4\textwidth}
   \centering
   \includegraphics[scale=0.35]{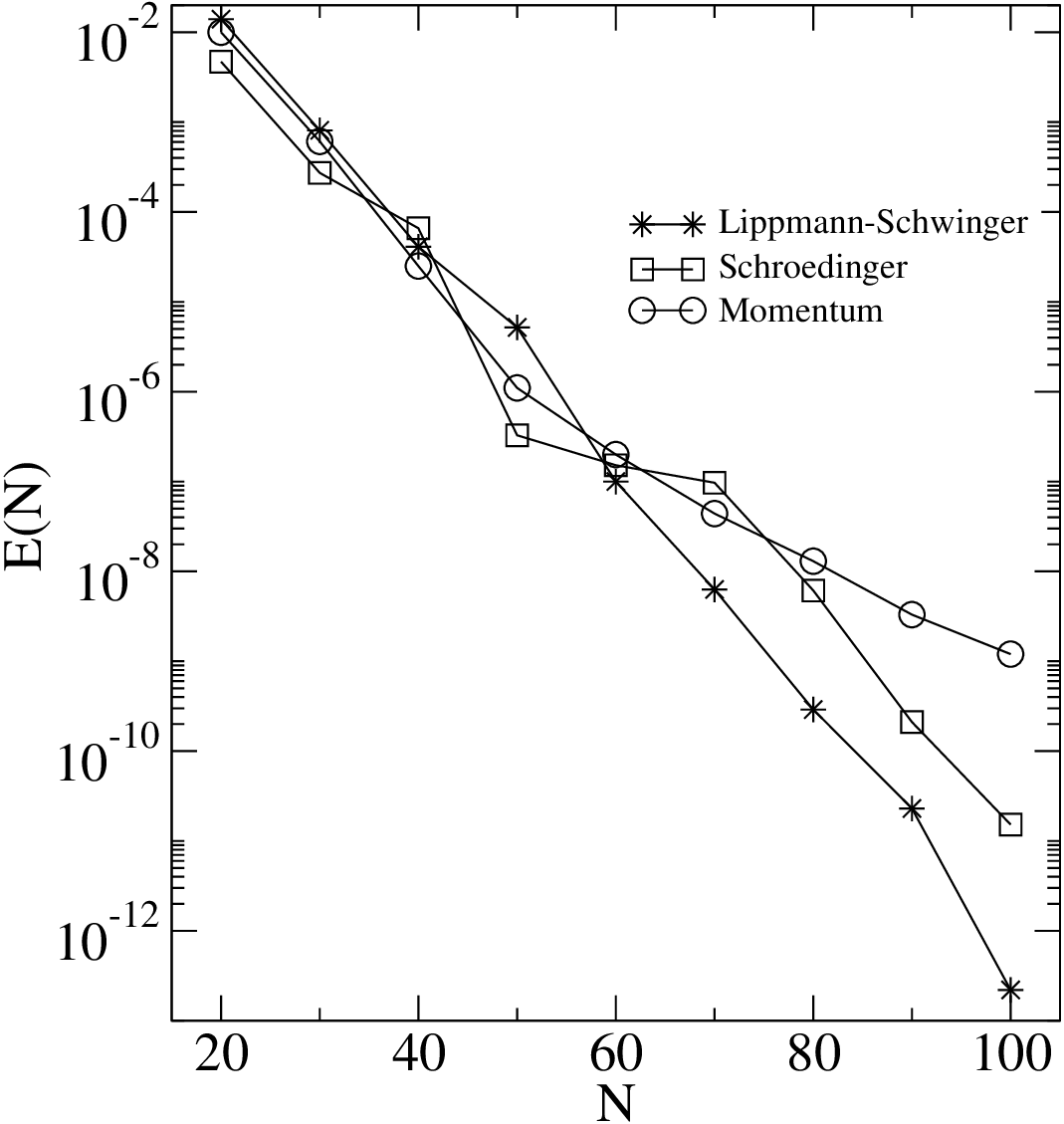}
   \caption{
   Average relative error (\ref{result1})
   for the  phase-shift
   versus $N$  (Morse potential).}
  \label{fig:four}
  \end{minipage}
  \vspace*{1.5cm}
  \end{figure}

 We begin with the continuous spectrum where we shall calculate the 
 s-wave phase-shift for three different potentials for which exact expressions
 are available (cf. Appendix). As a result of scaling, the phase-shift becomes
 a function of two dimensionless variables $\delta(s,\xi)$ where $\xi$ is
 the center-of-mass momentum in
 inverse range units ($\xi=pa$) and $s$ is the potential strength parameter.
 Since in each case we do know the exact phase-shift,
 the quantity of interest for us here will be the relative error. Our intention was 
 to remain close to the realm of nuclear physics and the strength parameter $s$ has been 
 chosen in such a way that the potentials roughly reproduce the  
 proton-proton $^{1}S_0$ phase-shift. This interaction is quite strong, being
 almost capable of supporting a bound state. The dependence upon $s$ will be examined later on.
 For each of the considered here potentials, keeping $s$ fixed,    
 we have computed the phase-shifts at $n_p$ equidistant values of 
 the momentum $\xi_i;\;i=1,2,\cdots n_p$ and this has been done repeatedly for 
 gradually increasing approximation order $N$. For an assigned approximation order $N$,
 we may define an average relative error $E(N)$, given as
 \begin{equation}
         E(N) = \dfrac{1}{n_p}\sum_{i=1}^{n_p}\dfrac{|\delta(s,\xi_i)-\delta_N(s,\xi_i)|}
         {|\delta(s,\xi_i)|},
         \label{result1}
 \end{equation}
 where $\delta$ and $\delta_N$ denote, respectively, the exact and the 
 N-th approximant to the computed phase-shift. In the actual computations
 we took $n_p=100$ with $s=0.8$ for both, exponential and Hulth\'{e}n potential
 and $s=0.2$ for Morse potential. The corresponding phase
 shifts vs. the center-of-mass momentum  $\xi$ are presented in fig. \ref{fig:one}.
 Among the considered potentials, 
 only the Morse potential, which is repulsive at small
 separations, is capable of producing a cross-over in the phase-shift.
 The error defined in (\ref{result1})  depends upon the order of the semi-spectral approximation
 $N$ used in the computation of $\delta_N$ and the quality of the approximation will be
 determined by the rate at which the error goes to zero for large $N$.
 The function $E(N)$ reflecting   
 the convergence of the semi-spectral method is displayed in 
 figs. \ref{fig:two} - \ref{fig:four} for the three different potentials. 
 \par
\begin{figure}
  \vspace*{1cm}
 \centering      
 \begin{minipage}[t]{.4\textwidth}
 \centering
  \includegraphics[scale=0.35]{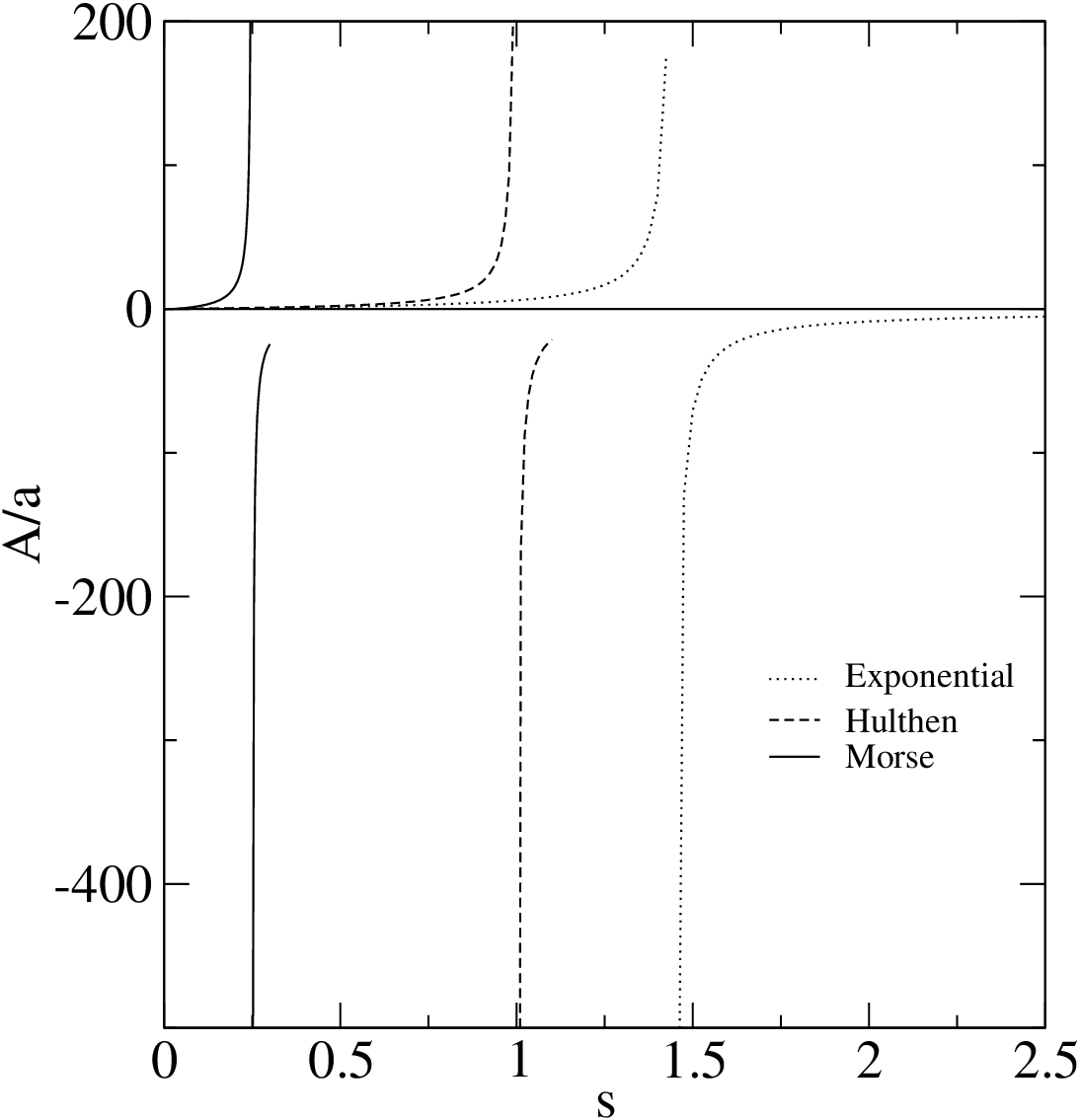}
       \caption{Scattering length 
        in inverse range units vs. $s$
           for different potentials. }
     \label{fig:five}
   \end{minipage}
  \hspace{1cm}
   \begin{minipage}[t]{.4\textwidth}
   \centering
   \includegraphics[scale=0.35]{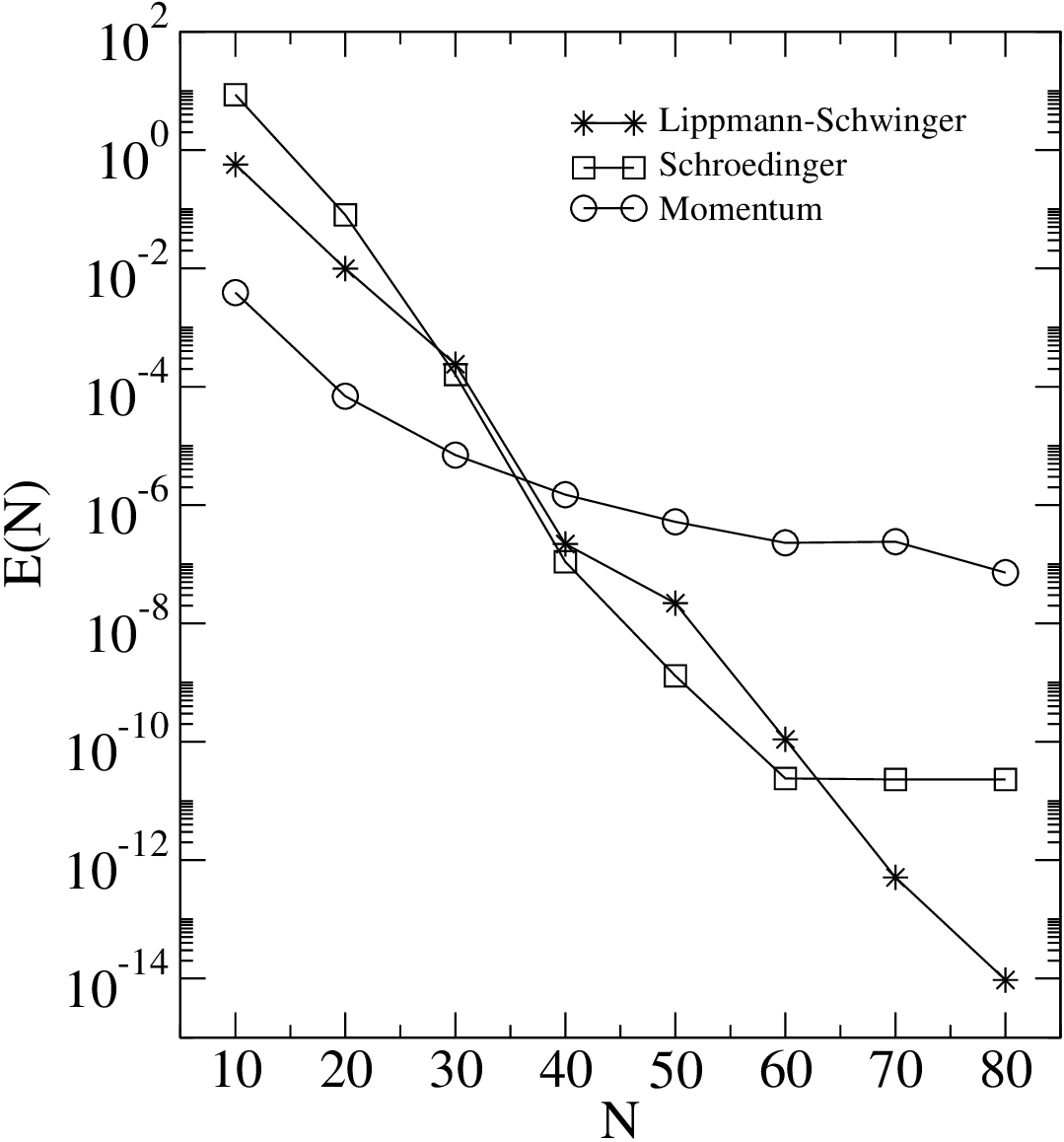}
   \caption{
   Average relative error (\ref{result2})
   for the scattering length
   versus $N$ (exponential potential).}
  \label{fig:six}
  \end{minipage}
 \vspace*{1.5cm}
  \end{figure}
\begin{figure}
 \vspace*{1cm}
 \centering      
 \begin{minipage}[b]{.4\textwidth}
 \centering 
  \includegraphics[scale=0.35]{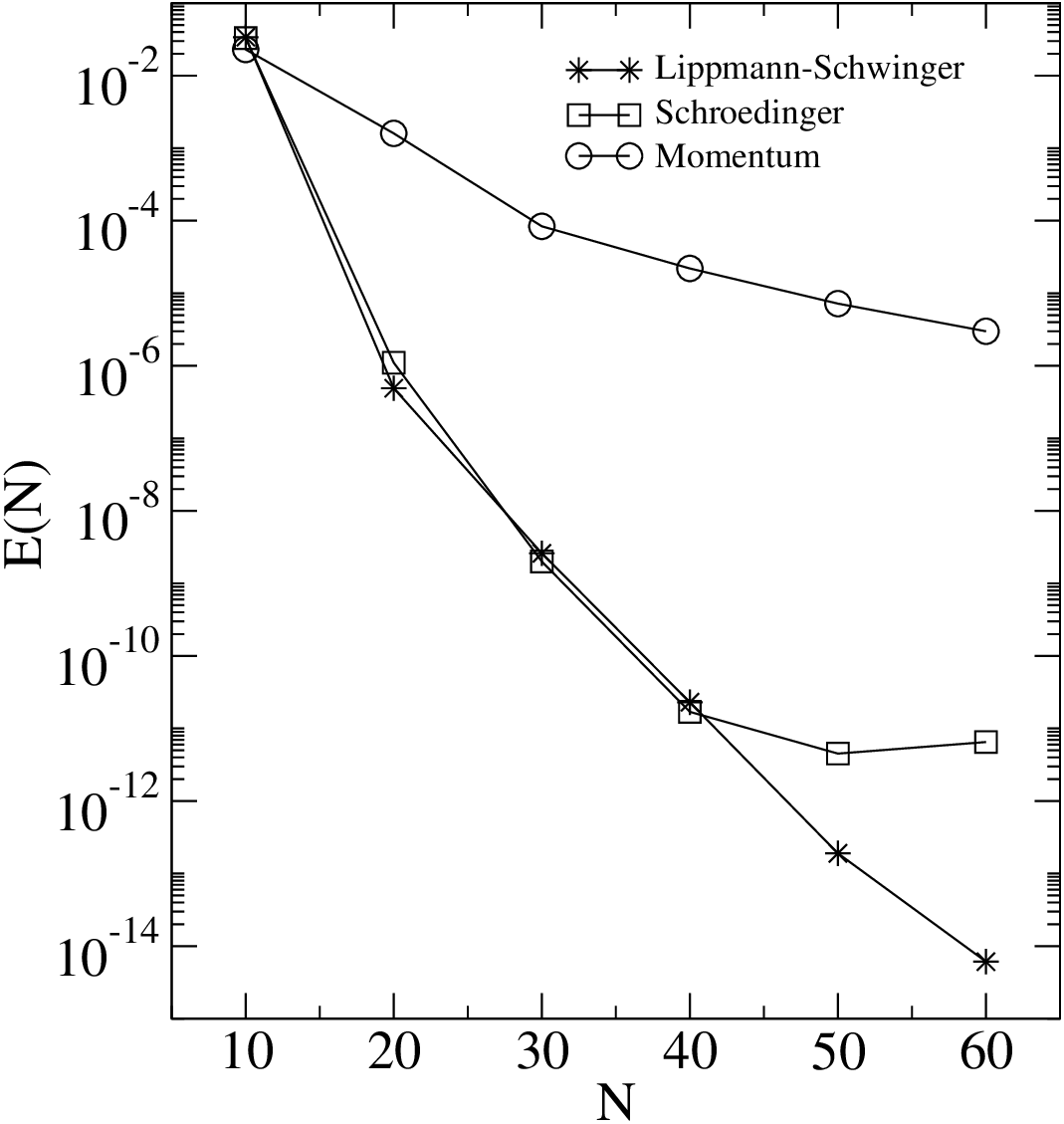}
   \caption{
   Average relative error (\ref{result2})
   for the scattering length
   versus $N$ (Hulth\'{e}n potential).}
     \label{fig:seven}
   \end{minipage}
   \hspace{1cm}
   \begin{minipage}[b]{.4\textwidth}
   \centering
   \includegraphics[scale=0.35]{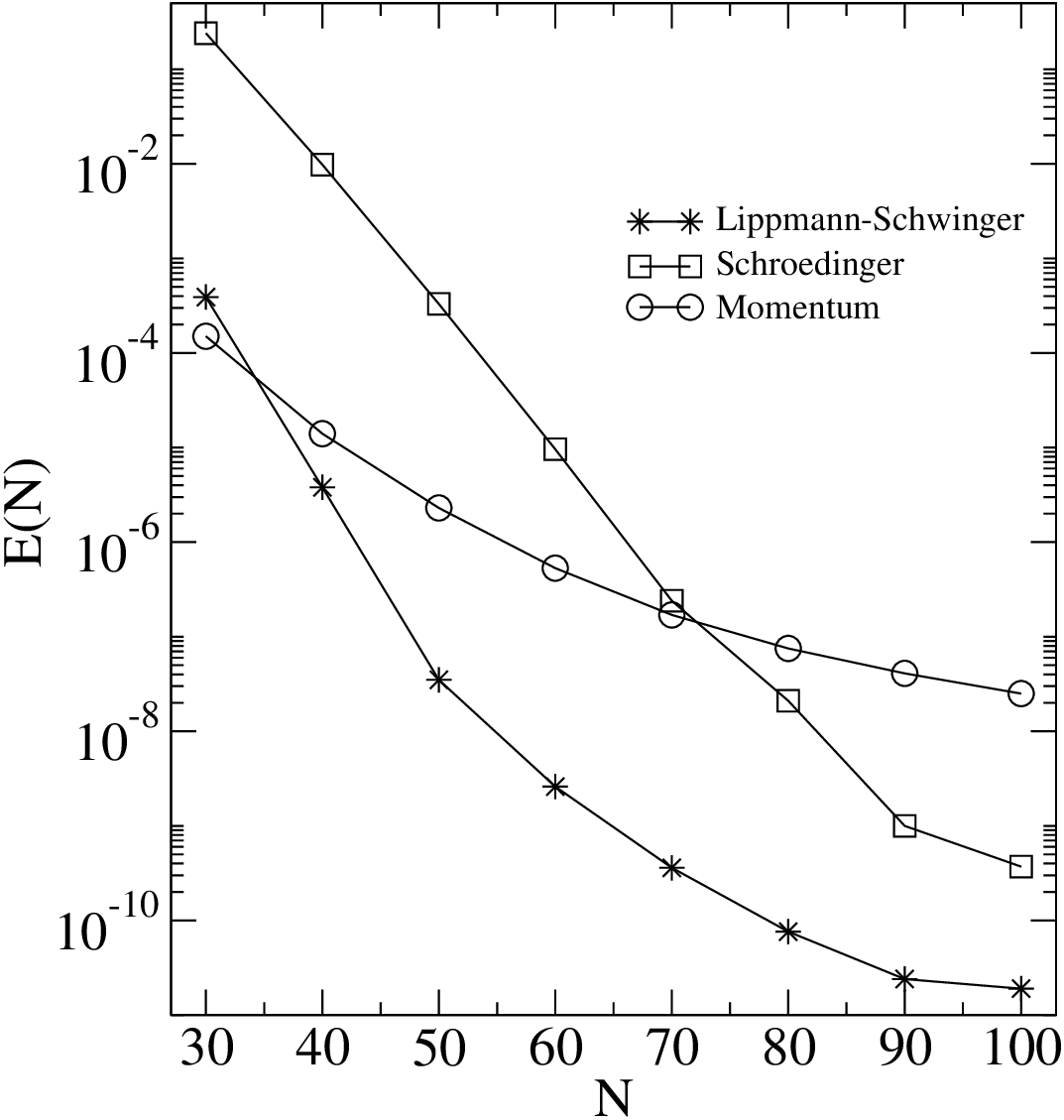}
   \caption{
   Average relative error (\ref{result2})
   for the scattering length
   versus $N$ (Morse potential).}
  \label{fig:eight}
  \end{minipage}
  \vspace*{1.5cm}
  \end{figure}

 In the procedure described above the value of the potential strength was
 frozen while we were calculating the phase-shifts at the different momenta.
 Now, we wish to examine the opposite situation when the momentum is fixed and
 the strength is allowed to vary. However, instead of the phase-shift
 we shall consider the scattering length which is known to posses a complicated structure
 as a function of $s$. This function exhibits interlacing zeros and poles
 and changes by many order of magnitude for $s$ values close to a pole, assuming
 both, positive and negative values. The plots of the scattering length vs. $s$
 for different potentials are presented in fig. \ref{fig:five}.  
 Similarly as before, we consider $n_p$ values of the strength $s_i;\;i=1,2,\dots n_p$
 and introduce the error function
 \begin{equation}
         E(N) = \dfrac{1}{n_p}\sum_{i=1}^{n_p}\dfrac{|(A(s_i)-A_N(s_i)|}
         {|A(s_i)|},
         \label{result2}
 \end{equation}
  where $A(s_i)$ and $A_N(s_i)$ denote, respectively, the exact scattering length
  (cf. Appendix) and its N-th approximant computed by the semi-spectral method.
  In (\ref{result2}) we just take the arithmetic mean of the relative errors
  computed for different $s$.  In the actual computations we took
 $n_p=100$  adopting for  $s_i$ an equidistant sequence of values in the
 intervals  $(0,\,2.5)$, $(0,\,1.1)$, $(0,\,0.3)$ for the exponential,
 Hulten and Morse potential, respectively.  As seen from fig. \ref{fig:five},
 in each case   
 the considered interval contained the $s$ value at which the scattering length
 exhibits a pole.  These critical values of the strength are
 $s\approx 1.44577$  for the exponential, 
 $s=1$ for the Hulth\'{e}n and $s= \frac{1}{4}$ for the Morse
 potential, respectively.  
  The error function obtained from (\ref{result2}) 
  for different potentials is given in figs. \ref{fig:six}-\ref{fig:eight}.
 \begin{figure}
  \vspace*{1cm}
 \centering      
 \begin{minipage}[t]{.4\textwidth}
 \centering
  \includegraphics[scale=0.35]{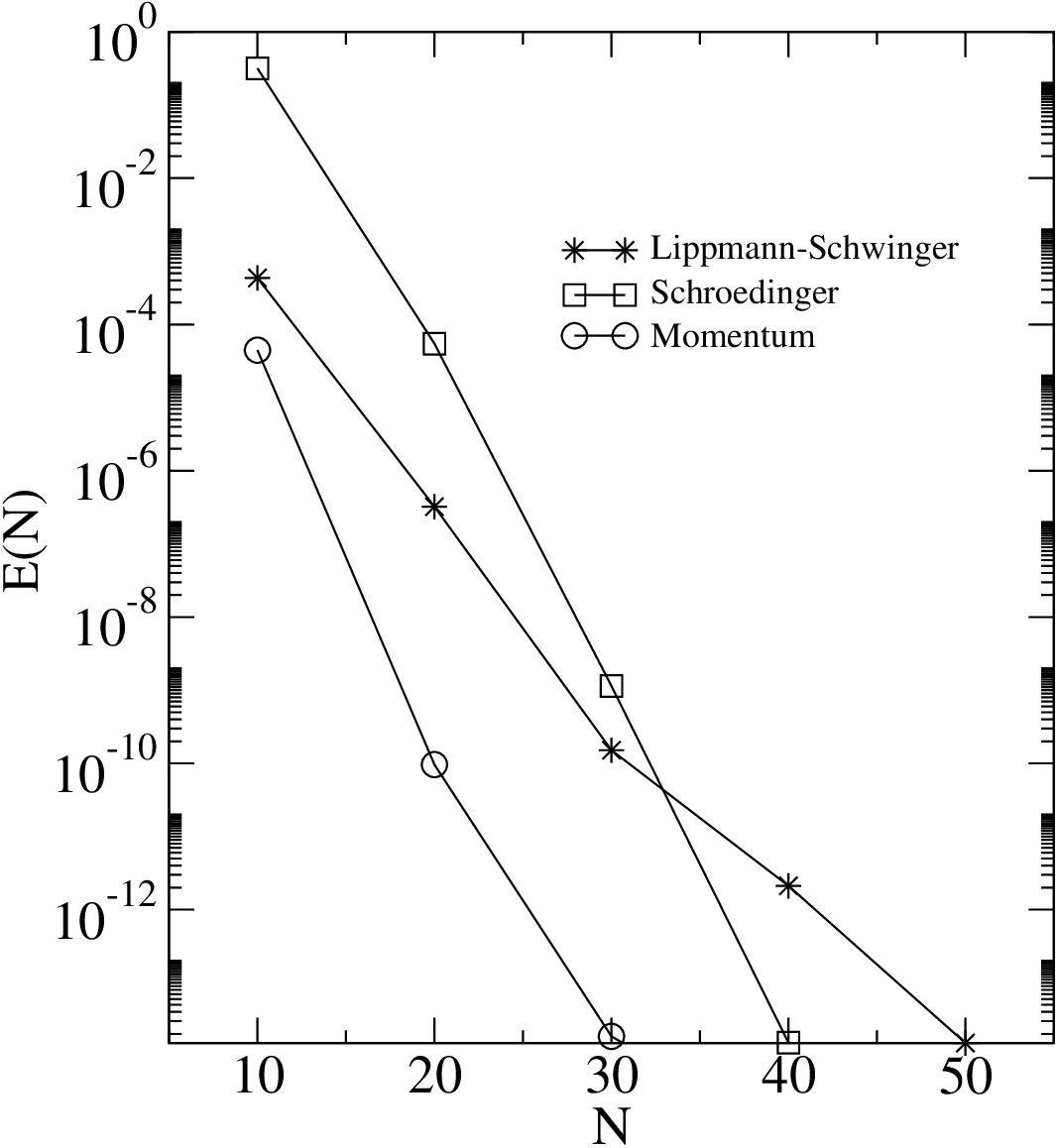}
   \caption{Relative error for the calculated
   binding energy -- exponential potential.}
     \label{fig:nine}
   \end{minipage}
  \hspace{1cm}
   \begin{minipage}[t]{.4\textwidth}
   \centering
   \includegraphics[scale=0.35]{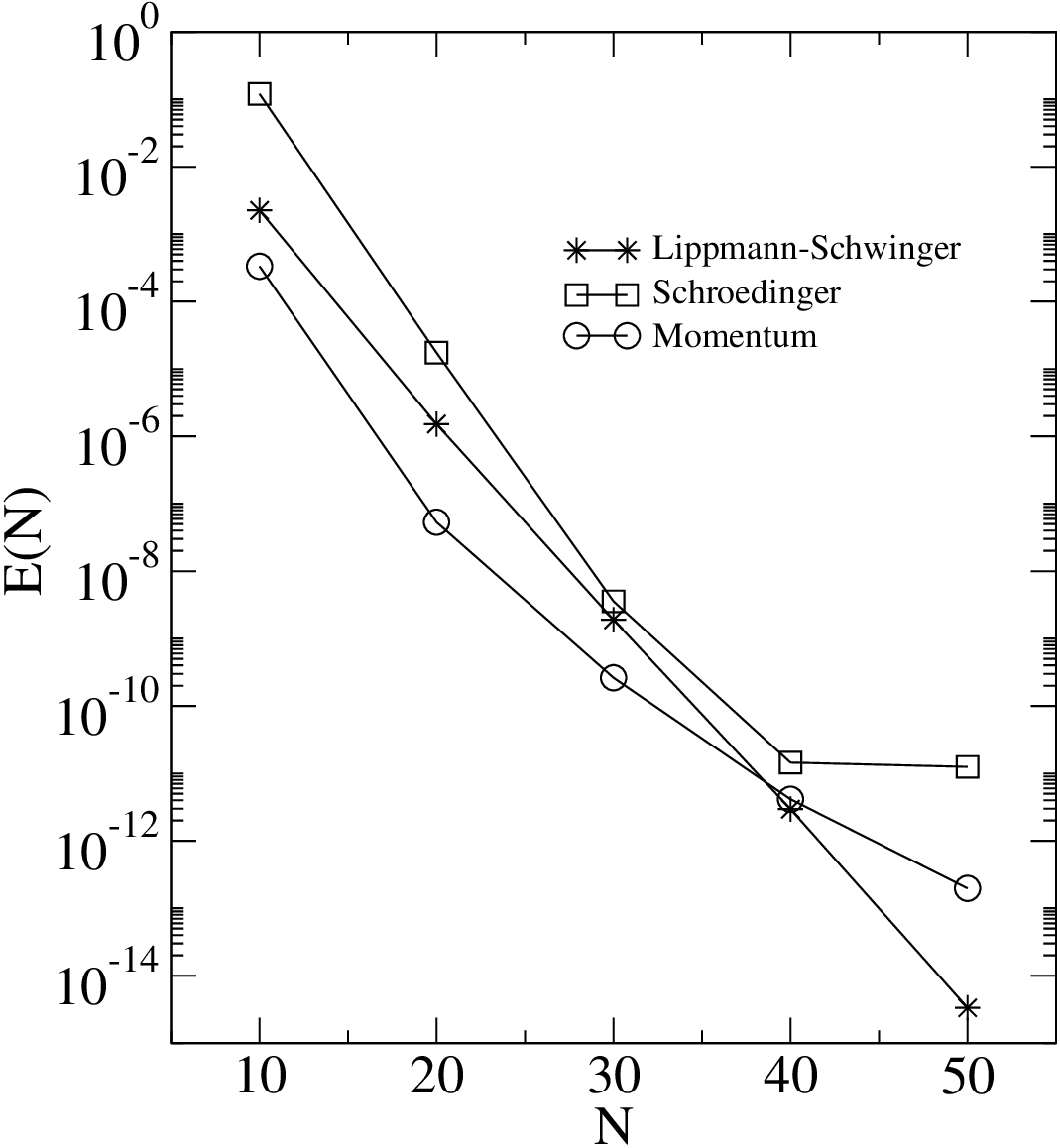}
   \caption{Relative error for the calculated
   binding energy -- Hulth\'{e}n potential.}
  \label{fig:ten}
  \end{minipage}
 \vspace*{1.5cm}
  \end{figure}
 \begin{figure}
\vspace*{1cm}
 \centering      
 \begin{minipage}[t]{.4\textwidth}
 \centering
  \includegraphics[scale=0.35]{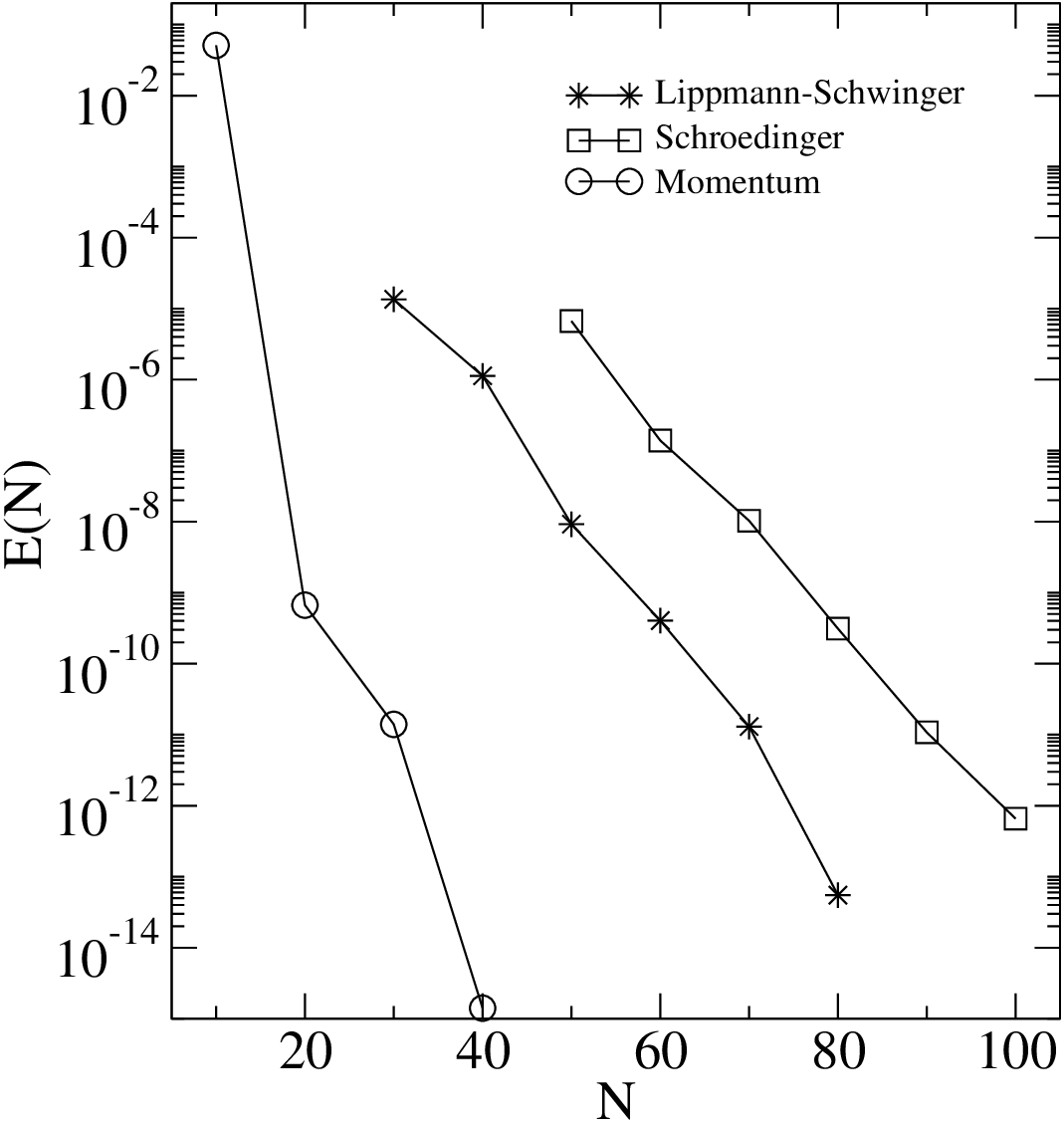}
   \caption{Relative error for the calculated
   binding energy -- Morse potential.}
     \label{fig:eleven}
   \end{minipage}
  \hspace{1cm}
   \begin{minipage}[t]{.4\textwidth}
   \centering
   \includegraphics[scale=0.35]{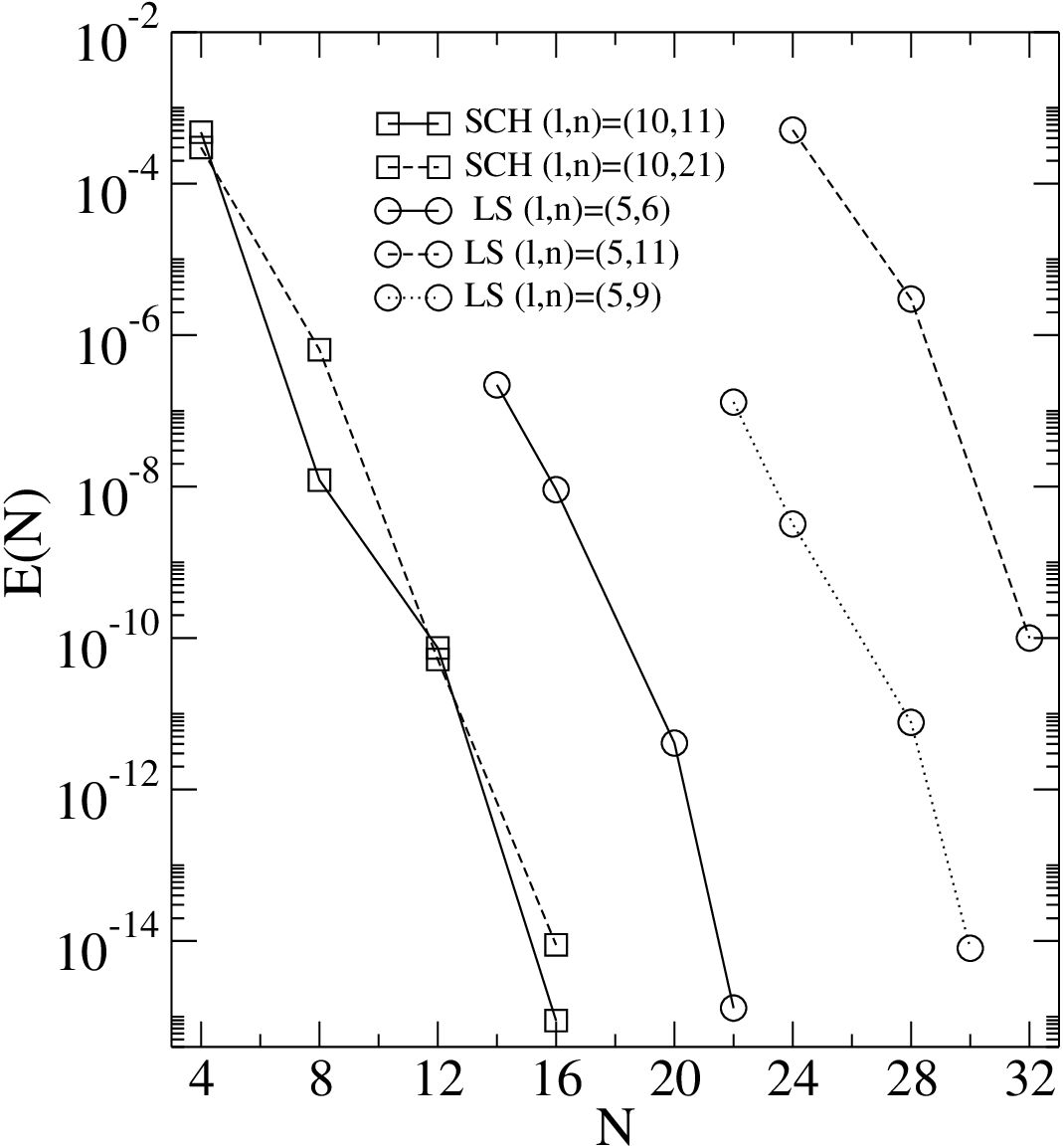}
   \caption{Relative error for the calculated
   binding energies -- Coulomb potential.}
  \label{fig:twelve}
  \end{minipage}
  \end{figure}
 In the discrete spectrum problem the dimensionless quantity of interest is
 the imaginary part of the momentum (in inverse range units) at which the T-matrix
 has a pole. This, necessarily non-negative quantity, will be denoted as $x(s)$ where 
 we have emphasised  its dependence upon the potential strength $s$. 
 Thus, similarly as for the scattering length $A(s)$, a function of $s$ needs to be
 determined but it is much harder to obtain the exact solution.
 The function $x(s)$ is known explicitly for the Hulth\'{e}n potential 
 but (cf. Appendix) for the exponential and the Morse potential
 $x(s)$ is provided in an entangled form, as a solution of a transcendental equation
 of the form $F(x,s)=0$ where the function $F$ might be quite complicated.
 In general, such equation can be solved only numerically and this is a difficult
 task especially when machine accuracy is desired. Nevertheless,
 for the exponential potential for some particular $x$ values, $s(x)$ 
 can be obtained quite easily and one such obvious instance is   
 $x=\Quarter$ in which case we get $s=(\pi/2)^2$ as the
 exact solution. Unfortunately, for the Morse potential there is only the
 hard way left. For testing purposes of the semi-spectral method as the 
 benchmark solutions we used the exact values $s=(\pi/2)^2,\; x=\Quarter$
 for exponential potential and, respectively,  $s=3/2,\;x=1$ for the Hulth\'{e}n potential.
 For the Morse potential the benchmark solution was obtained numerically.  
 The shape was adopted from \cite{darewychgreen} where Morse potential was used
 as a model of proton-neutron interaction in the deuteron
 (with $a=0.3408\, fm$ and $d=0.8668\,fm$). We slightly retouched the depth
 so that the potential better reproduces the deuteron binding energy $B=-2.22\,MeV$
 and at the pole, we have $s=0.33509414149514,\; x=0.0078864302204068$.
 The above pairs of values of $(s,x)$ for different potentials were regarded as 
 exact for testing purposes of the semi-spectral method.
 Using the $s$ value as input, the values of $x$ was subsequently determined 
 by solving the Schr\"{o}dinger equation, and the L-S equation in the coordinate
 and in momentum space, respectively. The relative error for $x$ is presented
 in Figs. \ref{fig:nine}-\ref{fig:eleven} as a function of the approximation order $N$.
 Unexpectedly, a comparison of the results obtained by different methods in 
 Figs. \ref{fig:two}-\ref{fig:eleven} shows that
 the unpopular Volterra equation method appears to be the winner.
 \par
 Finally, in Fig. \ref{fig:twelve} we display the relative errors for the calculated
 binding energies in a hydrogen-like system where a point-like charge Coulomb
 potential was operative. The values of the quantum numbers $(n,\,\ell)$ are
 given on the plot. We wish to note that in the momentum space calculation 
 there was no need to invoke Lande's subtraction technique and the difficulty
 connected with the Coulomb singularity was avoided by using the dedicated weights
 $\Omega_j(z)$, as explained in detail in Sec. VII. Therefore, the results presented in
 Fig. \ref{fig:twelve} indirectly involve also a test on the accuracy of the 
 method applied for calculating the singular integrals. 
 \par
  We have been displaying the logarithm of the error
  versus $N$ as this is convenient for a quick assessment of the asymptotic  
  convergence rate and a linear plot would indicate a geometric convergence
  (as a reminder we note that an algebraic convergence would lead to a linear
  dependence between $\log(E)$ and $\log(N)$).
  It has to be kept in mind, however, that
  with $N\leq 100$, as adopted in our computations,
  the approximation order $N$
  cannot be really regarded as asymptotic ($N\to\infty$)
  and a more complicated dependence upon $N$ should not be surprising.  
  Nevertheless, as might have
  been expected,  the semi-spectral errors presented in this Section
  decrease very rapidly to zero,
  in most cases showing indeed a geometric convergence.
  \section{Summary and conclusions}
  The purpose of this work was a concise presentation 
  of the mathematical techniques associated with the Chebyshev 
  pseudo-spectral method. The emphasis was on the practical aspects in the  
  efficient implementation of this method on high-performance computers,  
  in a manner accessible to newcomers to the field. 
  The paper is self-contained, i.e. 
  apart from a standard linear algebra package, no other resources are needed
  and the supplied algorithms
  can be immediately turned into a suite of computer codes capable of 
  solving a wide variety of practical problems.
  Formal considerations have been illustrated by providing
  concrete solutions of some typical quantum mechanical problems.  
  \par
  In our experience the pseudo-spectral Chebyshev method 
  seems to be without downsides: the implementation and programing 
  could not have been simpler, 
  the precision and stability are excellent with the
  convergence being usually exponential. Although the resulting matrices are
  not sparse but it does not matter since their sizes are not large.
  Finally,  the same scheme can be
  used for solving  differential, integral or integro-differential equations.  
 \appendix 
 \section{Configuration space}
 Denoting the range as $a$ and the depth as $V_0$, 
 we introduce two dimensionless quantities: the 
 potential strength parameter $s=2\mu\, V_0\, a^2$ 
 and the momentum in inverse range units
 $\xi=pa$.  The s-wave phase shift $\delta(s,\xi)$ is a function 
 of $s$ and $\xi$
 and the scattering length in the units of range $A(s)/a$ 
 is defined as the limit of $\delta(s,\xi)/\xi$ when $\xi\to 0$.
 \begin{enumerate}[(a)]
         \item{\it Exponential potential}.
 \begin{displaymath}
        2\mu\, V(r) = -( s/a^2)\,\E^{\displaystyle -r/a}
 \end{displaymath}
 and only the attractive case will be considered $s\geq 0$. 
 The phase shift, is
 \begin{displaymath}
 \delta(s,\xi)=\Im\ln\left[J_{2\I \xi}(2\sqrt{s})\,\Gamma(1+2\I \xi)\right] - \xi\,\ln s
 \end{displaymath}
 where $J_\nu(x)$ denotes a complex order $\nu$ Bessel function of the first kind.
 The scattering length, is
 \begin{displaymath}
         A(s)/a= 
         -2(\gamma+\log\sqrt{s})+\pi\,Y_0(2\sqrt{s})/J_0(2\sqrt{s}),
 \end{displaymath}
 where $\gamma$ is Euler constant and $Y_0(x)$ is the 
 zero-order Bessel function of the second  kind.
 For an assigned value of $s$, the
 bound states  poles are at $\xi=\I\, x$ and are obtained 
 by solving the transcendental equation
 \begin{displaymath}
        J_{2x}(2\sqrt{s})=0
 \end{displaymath}
\item{\it Hulth\'{e}n potential}.
 \begin{displaymath}
        2\mu\, V(r) = -( s/a^2)\, \left( \E^{\displaystyle r/a}-1\right)^{-1}
 \end{displaymath}
 and only the attractive case will be considered $s\geq 0$. 
 The phase shift, is 
 \begin{displaymath}
  \delta(s,\xi)=\Im\left[\ln\Gamma(1+2\I \xi)+\ln\Gamma(\sqrt{s-\xi^2}-\I \xi)   
  +\ln\Gamma(-\sqrt{s-\xi^2}-\I \xi)\right]      
 \end{displaymath}
 and the scattering length, is
 \begin{displaymath}
         A(s)/a=
    2\gamma -\psi(1+\sqrt{s})-\psi(1-\sqrt{s}),
 \end{displaymath}
 where $\psi$ denotes the digamma function.
 The bound state poles occur at $\xi=\I x$ where
 \begin{displaymath}
        x = (s-n^2)/(2n),\quad n=1,2,3\dots
 \end{displaymath}
\item{\it Morse potential} and {\it Morse barrier}.\\
 In both cases the potential is given by the same expression	
 \begin{displaymath}
        2\mu\, V(r) = -( s/a^2)\,\E^{\displaystyle (d-r)/a}
        \left[2-\E^{\displaystyle (d-r)/a} \right]
 \end{displaymath}
 where $d$ is a parameter of the dimension of length.
 For Morse potential we have $s>0$, whereas $s<0$ for Morse barrier.
 The phase shift, is
 \begin{displaymath}
     \delta(s,\xi)=
     \begin{cases} 	
	\arg M(\Half+\I \displaystyle \xi 
        -\sqrt{s},\,1+2\I \xi,\, z), & \quad s>0\\      
        \arg \,M(\Half+\I \displaystyle \xi 
        -\I\sqrt{-s},\,1+2\I \xi,\,\I\,z)       
        -\Half\,z, & \quad s\leq 0 
     \end{cases} 	
 \end{displaymath}
 where $ z=2 \E^{\displaystyle d/a}\, \sqrt{|s|} $
 and $M(a,b,z)$ denotes the Kummer function \cite{abramowitz}.
 The scattering length, takes the form 
 \begin{displaymath}
	 A(s)/a=
         -2\gamma-\psi(\Half-\sqrt{s})-\log z-\Gamma(\Half-\sqrt{s})\;
 U(\Half-\sqrt{s},\,1,\,z)/ M(\Half-\sqrt{s},\,1,\,z)
 \end{displaymath}
 for $s>0$ and
 \begin{displaymath}
	 A(s)/a=
         -2\gamma-\Re\left\{
         \psi(\Half-\I\sqrt{-s})-\log\I z+\Gamma(\Half-\I\sqrt{-s})\;
     U(\Half-\I\sqrt{-s},\,1,\,\I z)/M(\Half-\I\sqrt{-s},\,1,\,\I z) \right\}
 \end{displaymath}
 for $s\leq0$
 where $U(a,b,z)$ denotes the Tricomi hypergeometric function \cite{abramowitz}.
 For assigned values of $s$ and $d/a$, the bound state poles are
 at $\xi=\I x$ where $x$ is a solutions of the equation
 \begin{eqnarray*}
         M(\Half +x - \sqrt{s},\,1+2x,\, z)=0,\quad s>0\\
        \Re M(\Half +x - \I\sqrt{-s},\,1+2x,\, \I z) = 0,\quad  s<0. 
 \end{eqnarray*}
\item{\it Coulomb potential.}\\
For a Coulomb potential of the form $V(r)=-\alpha/r$ where $\alpha$ is the fine structure
constant, we have
\begin{equation*}
2\mu\,V(r) = -(2/a^2)\,(a/r)
\end{equation*} 
where  $a=1/\mu\alpha$ is the Bohr radius
 and only the attractive case will be considered. 
The phase shift, is
\begin{equation*}
\delta_\ell(\xi)=\arg\Gamma(\ell+1-\I/\xi).
\end{equation*} 
  Bound states occur at $\xi=\I\,x$ where
\begin{equation*}
x=1/(n+\ell+1),   \quad   n=0,1,2,\dots.
\end{equation*} 
When a $Z$ factor multiplies the Coulomb potential, as result of scaling
$r\to r/Z$ we get $x\to Z\,x$.
 \end{enumerate}
\section{Momentum space}
 The potentials in momentum space are obtained from (\ref{sing10}). 
 Introducing the abbreviations $x=a(k+k'),\;y=a(k-k')$, we have 
 \begin{enumerate}[(a)]
         \item{\it Exponential potential:}
 \begin{displaymath}
        U_0(k,k')=-2sa/[(1+x^2)(1+y^2)],
 \end{displaymath}
\item{\it Hulth\'{e}n potential:}
 \begin{displaymath}
        U_0(k,k')=-2sa\;\dfrac{\Re\psi(1+\I\,x)-\Re\psi(1+\I\,y)}{x^2-y^2}
        =-2sa\;\sum_{n=1}^\infty\dfrac{n}{(n^2+x^2)(n^2+y^2)},
 \end{displaymath}
\item{\it Morse potential and Morse barrier:}
 \begin{displaymath}
        U_{0}(k,k')=-sa\;\dfrac{4\,\E^{d/a}}{(1+x^2)(1+y^2)}
        +sa\;\dfrac{1}{4}\;\dfrac{\E^{2d/a}}{[1+(\Half x)^2][1+(\Half y)^2]},
 \end{displaymath}
\item{\it Coulomb potential:}
\begin{equation*}
 U_\ell(k,k')=-4a\; Q_\ell(z)/(x^2-y^2)  
\end{equation*} 
where $z=(x^2+y^2)/(x^2-y^2)$. 
 Following \cite{abramowitz},
 $Q_\ell(z)$ denotes the Legendre function of the second kind which exhibits 
 a logarithmic singularity 
\begin{displaymath}
        Q_\ell(z)=P_\ell(z)\;\Half\log\dfrac{1+z}{1-z}-W_{\ell-1}(z)
\end{displaymath}
with $W_{-1}(z)\equiv 0$ and
\begin{displaymath}
        W_{\ell-1}(z)=
 \displaystyle\sum_{n=1}^\ell\frac{1}{n}P_{n-1}(z)\,P_{\ell-n}(z),
\end{displaymath}
where $P_\ell(z)$ denotes the Legendre polynomial.
 \end{enumerate}

 \end{document}